\documentclass[prd,twocolumn,showpacs,amsmath,amssymb,superscriptaddress]{revtex4}

\usepackage{epsfig}
\usepackage{psfrag}
\usepackage{graphicx}
\usepackage{dcolumn}
\usepackage{bm}
\usepackage{epsfig}
\usepackage{multirow}
\usepackage{slashed}
\usepackage{rotating}
\usepackage{subfig}
\usepackage{caption}

\newcommand{\pt}{\mbox{$p_{\rm T}$}}

\newcommand{\etadet}{\mbox{$\eta_{\rm det}$}}

\newcommand{\mll}{\mbox{$m_{\ell \ell}$}}

\newcommand{\kzjets}{\mbox{$k_{Z+{\rm jets}}$}}
\newcommand{\dzero}{D0}

\newcommand{\ee}      {\ensuremath{ee}}
\newcommand{\mumu}    {\ensuremath{\mu\mu}}
\newcommand{\eeicr}   {\ensuremath{ee_{\rm ICR}}}
\newcommand{\mumutrk} {\ensuremath{\mu\mu_{\rm trk}}}

\newcommand{\mutrk}   {\ensuremath{\mu_{\rm trk}}}

\newcommand{\GeV} {\ensuremath{\mathrm{Ge\kern -0.1em V}}}
\newcommand{\TeV} {\ensuremath{\mathrm{Te\kern -0.1em V}}}
\newcommand{\ifb} {\ensuremath{{\rm fb}^{-1}}}

\newcommand{\ppbar}{\mbox{$p\overline{p}$}}

\newcommand{\ttbar}{\mbox{$t\overline{t}$}}

\newcommand{\alpgen}{{\sc alpgen}}
\newcommand{\pythia}{{\sc pythia}}
\newcommand{\mcfm}{{\sc mcfm}}
\newcommand{\geant}{{\sc geant3}}

\begin{document}




\hspace{5.2in} \mbox{FERMILAB-PUB-10-325-E}

\title{\boldmath 
Search for $ZH \rightarrow \ell^+\ell^-b\bar{b}$ production in
$4.2$~fb$^{-1}$ of $p\bar{p}$ collisions at $\sqrt{s}=1.96~\TeV$}



%
\affiliation{Universidad de Buenos Aires, Buenos Aires, Argentina}
\affiliation{LAFEX, Centro Brasileiro de Pesquisas F{\'\i}sicas, Rio de Janeiro, Brazil}
\affiliation{Universidade do Estado do Rio de Janeiro, Rio de Janeiro, Brazil}
\affiliation{Universidade Federal do ABC, Santo Andr\'e, Brazil}
\affiliation{Instituto de F\'{\i}sica Te\'orica, Universidade Estadual Paulista, S\~ao Paulo, Brazil}
\affiliation{Simon Fraser University, Vancouver, British Columbia, and York University, Toronto, Ontario, Canada}
\affiliation{University of Science and Technology of China, Hefei, People's Republic of China}
\affiliation{Universidad de los Andes, Bogot\'{a}, Colombia}
\affiliation{Charles University, Faculty of Mathematics and Physics, Center for Particle Physics, Prague, Czech Republic}
\affiliation{Czech Technical University in Prague, Prague, Czech Republic}
\affiliation{Center for Particle Physics, Institute of Physics, Academy of Sciences of the Czech Republic, Prague, Czech Republic}
\affiliation{Universidad San Francisco de Quito, Quito, Ecuador}
\affiliation{LPC, Universit\'e Blaise Pascal, CNRS/IN2P3, Clermont, France}
\affiliation{LPSC, Universit\'e Joseph Fourier Grenoble 1, CNRS/IN2P3, Institut National Polytechnique de Grenoble, Grenoble, France}
\affiliation{CPPM, Aix-Marseille Universit\'e, CNRS/IN2P3, Marseille, France}
\affiliation{LAL, Universit\'e Paris-Sud, CNRS/IN2P3, Orsay, France}
\affiliation{LPNHE, Universit\'es Paris VI and VII, CNRS/IN2P3, Paris, France}
\affiliation{CEA, Irfu, SPP, Saclay, France}
\affiliation{IPHC, Universit\'e de Strasbourg, CNRS/IN2P3, Strasbourg, France}
\affiliation{IPNL, Universit\'e Lyon 1, CNRS/IN2P3, Villeurbanne, France and Universit\'e de Lyon, Lyon, France}
\affiliation{III. Physikalisches Institut A, RWTH Aachen University, Aachen, Germany}
\affiliation{Physikalisches Institut, Universit{\"a}t Freiburg, Freiburg, Germany}
\affiliation{II. Physikalisches Institut, Georg-August-Universit{\"a}t G\"ottingen, G\"ottingen, Germany}
\affiliation{Institut f{\"u}r Physik, Universit{\"a}t Mainz, Mainz, Germany}
\affiliation{Ludwig-Maximilians-Universit{\"a}t M{\"u}nchen, M{\"u}nchen, Germany}
\affiliation{Fachbereich Physik, Bergische  Universit{\"a}t Wuppertal, Wuppertal, Germany}
\affiliation{Panjab University, Chandigarh, India}
\affiliation{Delhi University, Delhi, India}
\affiliation{Tata Institute of Fundamental Research, Mumbai, India}
\affiliation{University College Dublin, Dublin, Ireland}
\affiliation{Korea Detector Laboratory, Korea University, Seoul, Korea}
\affiliation{CINVESTAV, Mexico City, Mexico}
\affiliation{FOM-Institute NIKHEF and University of Amsterdam/NIKHEF, Amsterdam, The Netherlands}
\affiliation{Radboud University Nijmegen/NIKHEF, Nijmegen, The Netherlands}
\affiliation{Joint Institute for Nuclear Research, Dubna, Russia}
\affiliation{Institute for Theoretical and Experimental Physics, Moscow, Russia}
\affiliation{Moscow State University, Moscow, Russia}
\affiliation{Institute for High Energy Physics, Protvino, Russia}
\affiliation{Petersburg Nuclear Physics Institute, St. Petersburg, Russia}
\affiliation{Stockholm University, Stockholm and Uppsala University, Uppsala, Sweden }
\affiliation{Lancaster University, Lancaster LA1 4YB, United Kingdom}
\affiliation{Imperial College London, London SW7 2AZ, United Kingdom}
\affiliation{The University of Manchester, Manchester M13 9PL, United Kingdom}
\affiliation{University of Arizona, Tucson, Arizona 85721, USA}
\affiliation{University of California Riverside, Riverside, California 92521, USA}
\affiliation{Florida State University, Tallahassee, Florida 32306, USA}
\affiliation{Fermi National Accelerator Laboratory, Batavia, Illinois 60510, USA}
\affiliation{University of Illinois at Chicago, Chicago, Illinois 60607, USA}
\affiliation{Northern Illinois University, DeKalb, Illinois 60115, USA}
\affiliation{Northwestern University, Evanston, Illinois 60208, USA}
\affiliation{Indiana University, Bloomington, Indiana 47405, USA}
\affiliation{Purdue University Calumet, Hammond, Indiana 46323, USA}
\affiliation{University of Notre Dame, Notre Dame, Indiana 46556, USA}
\affiliation{Iowa State University, Ames, Iowa 50011, USA}
\affiliation{University of Kansas, Lawrence, Kansas 66045, USA}
\affiliation{Kansas State University, Manhattan, Kansas 66506, USA}
\affiliation{Louisiana Tech University, Ruston, Louisiana 71272, USA}
\affiliation{University of Maryland, College Park, Maryland 20742, USA}
\affiliation{Boston University, Boston, Massachusetts 02215, USA}
\affiliation{Northeastern University, Boston, Massachusetts 02115, USA}
\affiliation{University of Michigan, Ann Arbor, Michigan 48109, USA}
\affiliation{Michigan State University, East Lansing, Michigan 48824, USA}
\affiliation{University of Mississippi, University, Mississippi 38677, USA}
\affiliation{University of Nebraska, Lincoln, Nebraska 68588, USA}
\affiliation{Rutgers University, Piscataway, New Jersey 08855, USA}
\affiliation{Princeton University, Princeton, New Jersey 08544, USA}
\affiliation{State University of New York, Buffalo, New York 14260, USA}
\affiliation{Columbia University, New York, New York 10027, USA}
\affiliation{University of Rochester, Rochester, New York 14627, USA}
\affiliation{State University of New York, Stony Brook, New York 11794, USA}
\affiliation{Brookhaven National Laboratory, Upton, New York 11973, USA}
\affiliation{Langston University, Langston, Oklahoma 73050, USA}
\affiliation{University of Oklahoma, Norman, Oklahoma 73019, USA}
\affiliation{Oklahoma State University, Stillwater, Oklahoma 74078, USA}
\affiliation{Brown University, Providence, Rhode Island 02912, USA}
\affiliation{University of Texas, Arlington, Texas 76019, USA}
\affiliation{Southern Methodist University, Dallas, Texas 75275, USA}
\affiliation{Rice University, Houston, Texas 77005, USA}
\affiliation{University of Virginia, Charlottesville, Virginia 22901, USA}
\affiliation{University of Washington, Seattle, Washington 98195, USA}
\author{V.M.~Abazov} \affiliation{Joint Institute for Nuclear Research, Dubna, Russia}
\author{B.~Abbott} \affiliation{University of Oklahoma, Norman, Oklahoma 73019, USA}
\author{M.~Abolins} \affiliation{Michigan State University, East Lansing, Michigan 48824, USA}
\author{B.S.~Acharya} \affiliation{Tata Institute of Fundamental Research, Mumbai, India}
\author{M.~Adams} \affiliation{University of Illinois at Chicago, Chicago, Illinois 60607, USA}
\author{T.~Adams} \affiliation{Florida State University, Tallahassee, Florida 32306, USA}
\author{G.D.~Alexeev} \affiliation{Joint Institute for Nuclear Research, Dubna, Russia}
\author{G.~Alkhazov} \affiliation{Petersburg Nuclear Physics Institute, St. Petersburg, Russia}
\author{A.~Alton$^{a}$} \affiliation{University of Michigan, Ann Arbor, Michigan 48109, USA}
\author{G.~Alverson} \affiliation{Northeastern University, Boston, Massachusetts 02115, USA}
% ...
%
%
\collaboration{The D0 Collaboration} \noaffiliation
\vskip 0.25cm


\date{August 20, 2010}

\begin{abstract}
\vspace*{3.0cm} We present a search for the standard model Higgs boson 
produced in association with a $Z$ boson in 4.2 fb$^{-1}$ of $p\bar{p}$
collisions, collected with the \dzero\ detector at the Fermilab Tevatron at
$\sqrt{s}$ = 1.96 $\TeV$.  Selected events contain one reconstructed
$Z\rightarrow \ell^+\ell^-$ candidate and at least two jets, including at
least one $b$-tagged jet.  In the absence of an excess over the background
expected from other standard model processes, limits on the $ZH$ cross
section multiplied by the branching ratios are set.  The limit at $M_H =
115~\GeV$ is a factor of 5.9 larger than the standard model prediction.

\end{abstract}

\maketitle

In the standard model (SM), the spontaneous breakdown of the electroweak
gauge symmetry generates masses for the $W$ and $Z$ bosons and produces a
scalar massive particle, the Higgs boson, which has so far eluded detection.
The discovery of the Higgs boson would top a remarkable list of
experimentally confirmed SM predictions.  

For Higgs boson masses $M_H$ below $135~\GeV$, the primary Higgs boson decay
in the SM is $H \rightarrow b\bar{b}$, which is challenging to discern
amidst copious $b\bar{b}$ production at the Tevatron $p\bar{p}$ collider.
Consequently, sensitivity to a low-mass Higgs boson is predominantly from
its production in association with a $W$ or $Z$ boson that decays to
leptons.

In this Letter, we present a search for $ZH \rightarrow
\ell^+\ell^-b\bar{b}$, where $\ell$ is either a muon or an electron.  The
searches for $ZH\to\nu\bar{\nu}b\bar{b}$ and $ZH\to\tau^+\tau^-b\bar{b}$ are
treated elsewhere~\cite{nunubb,higgstau}. For the $\ell^+\ell^-b\bar{b}$
final states, the \dzero~collaboration has previously used 0.45$~\ifb$ of
integrated luminosity to report a cross section upper limit at the 95\% CL
that was around 25 times larger than the SM prediction at
$M_H=115~\GeV$~\cite{prevzh}, and the CDF~collaboration used 2.7$~\ifb$~to
obtain a factor of around 8~\cite{cdfzh}.  

The data for this analysis were collected at the Fermilab Tevatron Collider
with the \dzero~detector \cite{d0det}.  After imposing data quality
requirements, the integrated luminosity is $4.2~\ifb$.  The selected events
were predominantly acquired by triggers that provide real-time
identification of electron and muon candidates, but to maximize acceptance,
events from all available triggers are considered.

The selection of signal-like events requires a primary $p\bar{p}$
interaction vertex (PV) that has at least three associated tracks and is
located within 60 cm of the center of the detector along the direction of
the beam.  Selected events must also contain a $Z$ boson candidate with a
dilepton invariant mass $60 < \mll < 150~\GeV$.  

The dimuon ($\mumu$) selection requires at least two muons matched to
central tracks with transverse momenta $\pt >10~\GeV$.  Combined tracking
and calorimeter isolation requirements are applied to the muon pair such
that one muon does not need to be isolated if the other is sufficiently well
isolated. For each muon track, the pseudorapidity $\etadet$, measured with
respect to the center of the detector, must satisfy $|\etadet| <
2$~\cite{coordinates}. At least one muon must have $|\etadet| < 1.5$ and
$\pt > 15~\GeV$. The distance of closest approach of each track to the PV in
the plane transverse to the beam direction, $d_{\rm PV}$, must be less than
$0.02~$cm for tracks with at least one hit in the silicon microstrip tracker
(SMT).  A track without SMT hits must have $d_{\rm PV} < 0.2$~cm, and its
$\pt$ is corrected through a constraint to the position of the PV.  An
additional dimuon selection, $\mumutrk$, requires one identified muon and
one isolated track ($\mutrk$) in the central tracking detector with
$\pt>20~\GeV$ and $|\etadet| < 2$, at least one hit in the SMT, and $d_{\rm
PV}<0.02~$cm~\cite{strauss}.  The $\mutrk$ must be separated in
pseudorapidity $\eta$ and azimuth $\phi$ by $\Delta \mathcal{R} =
\sqrt{(\Delta \eta)^2+(\Delta \phi)^2}>0.1$ from the other muon.
The $\mumutrk$ selection adds $10\%$ signal acceptance to the $\mumu$
selection, mainly from gaps in the muon detector. To reduce contamination
from cosmic rays, the tracks from both selections must not be back-to-back
in $\eta$ and $\phi$. The two muons must also have opposite charge.  

The dielectron ($\ee$) selection requires at least two electrons of $\pt >
15~\GeV$ identified by electromagnetic showers in the calorimeter.  Each
shower must be isolated from other energy depositions and have a shape
consistent with that expected of an electron.  At least one electron must be
identified in the central calorimeter (CC, $|\etadet|<1.1$), and a second
electron either in the CC or the end calorimeter (EC, $1.5<|\etadet|<2.5$).
The CC electrons must match central tracks or produce a pattern of hits in
the tracker consistent with that expected of an electron.  An additional
dielectron selection, $\eeicr$, requires exactly one electron from the CC or
EC, with a second electron identified as a narrow calorimeter cluster in the
inter-cryostat region (ICR, $1.1<|\etadet|<1.5$) with a matching track in
the central tracker~\cite{calpas}.  A neural network (NN$_{\rm ICR}$) is
used to differentiate ICR electrons from jets. The $\eeicr$ selection
requires an explicit single-electron trigger, and adds $17\%$ signal
acceptance to the $\ee$ selection.

Jets are reconstructed in the calorimeter using the iterative midpoint cone
algorithm \cite{runiicone} with a cone of radius 0.5.  The energy scale of
jets is corrected for detector response, the presence of noise and multiple
$p\bar{p}$ interactions, and energy deposited outside of the reconstructed
jet cone. At least two jets with $|\etadet| < 2.5$ are required, with the
leading jet of $\pt > 20~\GeV$ and additional jets of $\pt > 15~\GeV$.  
Both electrons in dielectron events are required to be isolated from any jet
by $\Delta \mathcal{R} > 0.5$. Likewise, jets must be separated by $\Delta
\mathcal{R} > 0.5$ from the $\mutrk$ candidate in the $\mumutrk$ channel,
but no such requirement is applied to the muon candidates in either dimuon
channel.  To reduce the impact from multiple $p\bar{p}$ interactions at high
instantaneous luminosities, jets must contain at least two tracks matched to
the PV.

To distinguish the decay $H \rightarrow b\bar{b}$ from background processes
involving light quarks and gluons, jets are identified as likely containing
$b$-quarks ($b$-tagged) if they pass loose or tight requirements on the
output of a neural network trained to separate $b$-jets from light
jets~\cite{bid}.  For $|\eta|<0.7$ and $\pt>45~\GeV$, the $b$-tagging
efficiency for $b$-jets and the misidentification rate of light jets are,
respectively, $74\%$ and $8.5\%$ for loose $b$-tags, and $48\%$ and $0.6\%$
for tight $b$-tags. Events with at least two loose $b$-tags are classified
as double-tagged (DT).  Events not in the DT sample that contain a single
tight $b$-tag are classified as single-tagged (ST).  The dijet $H\rightarrow
b\bar{b}$ candidate is composed of the two highest $\pt$ $b$-tagged jets in DT
events, and the $b$-tagged jet plus the highest $\pt$ non-$b$-tagged jet in
ST events.

The background from multijet events with jets misidentified as leptons is
estimated from control samples in the data.  For the $\mumu$ channel, the
multijet control sample contains events that fail the muon isolation
requirement but otherwise pass the event selection.  In the $\mumutrk$
multijet control sample, the $\mu$ and $\mutrk$ are required to have the
same charge.  For the $\ee$ channel, the electrons must fail isolation and
shower shape requirements.  The resulting trigger bias is corrected by
reweighting distributions in lepton $\pt$ and $\eta$ to match an unbiased
control sample.  Misidentified ICR electrons in the $\eeicr$ channel are
selected from a background region of the NN$_{\rm ICR}$ output.

The dominant background process is the production of a $Z$ boson in
association with jets, with the $Z$ boson decaying to dileptons ($Z+$jets).
The light-flavor component ($Z+$LF) includes jets from only light quarks
($uds$) or gluons.  The heavy-flavor component ($Z+$HF) includes
non-resonant $Z+b\bar{b}$ production, which has the same final state as the
signal, and $Z+c\bar{c}$.  The remaining backgrounds are from top quark pair
($t\bar{t}$) and diboson production.  We simulate $ZH \rightarrow
\ell^+\ell^-b\bar{b}$ and inclusive diboson production with
\pythia~\cite{pythia} and $Z+$jets and $t\bar{t} \rightarrow \ell^+\nu
b\ell^-\bar{\nu}\bar{b}$ processes with \alpgen~\cite{alpgen}, using the
{\sc CTEQ6L1}~\cite{cteq6} leading-order parton distribution functions
(PDFs).  The events generated with \alpgen~are input to \pythia~for parton
showering and hadronization, and can contain additional jets.  For these
events, we use a matching procedure to avoid double counting partons
produced by \alpgen~and those subsequently added by the showering in
\pythia~\cite{alpgen}.  All samples are processed using a detector
simulation program based on \geant~\cite{geant}, and the same off\-line
reconstruction algorithms used to process the data. Events from randomly
chosen beam crossings are overlaid on the simulated events to reproduce the
effect of multiple $\ppbar$ interactions and detector noise.

The cross section and branching ratio for the signal are taken from 
Refs.~\cite{zhxsec,hdecay}.  For the $t\bar{t}$ and diboson processes, the
cross sections are taken from \mcfm~\cite{mcfm}, calculated at
next-to-leading order (NLO).  The inclusive $Z$ boson cross section is
scaled to next-to-NLO \cite{dyxsec}, with additional NLO heavy-flavor
corrections calculated from \mcfm~applied to $Z+b\bar{b}$ and $Z+c\bar{c}$.

Corrections are applied to the simulated events to improve the modeling.
The simulated $\eeicr$, $\mumu$ and $\mumutrk$ events are weighted by
trigger efficiencies measured in data.  For the $\ee$ channel, no correction
is applied as the combination of lepton and jet triggers is nearly 100\%
efficient. Lepton identification efficiencies are corrected as a function of
$\etadet$ and $\phi$ of the lepton.  Jet energies are modified to reproduce
the resolution observed in data. Scale factors are applied to correct for
differences in jet reconstruction efficiency between data and simulation. To
model the $b$-tagged samples, simulated events are weighted by their probability
to satisfy the ST or DT criteria as measured in data.

The performance of the background model is evaluated in control samples with
negligible signal contributions that are obtained by applying only the
lepton selection requirements (inclusive) or all selection requirements
except $b$-tagging (pretag).  The simulated $Z$ boson events are reweighted
such that the $\pt$ distribution of the $Z$ boson is consistent with the
observed distribution \cite{zptrw}. To improve upon the {\alpgen} modeling
of $Z+$jets, motivated by a comparison with the {\sc sherpa}
generator~\cite{sherpa}, the pseudorapidities of the two jets with the
highest $\pt$, and the $\Delta \mathcal{R}$ between them are reweighted to
match the distributions measured in the pretag data.

Normalization factors for the simulated and the multijet samples are
determined from a fit to the $\mll$ distributions in the inclusive and
pretag data. This improves the accuracy of the background model and reduces
the impact of systematic uncertainties that affect pretag event yields
(e.g., uncertainties on luminosity). The region $40 < \mll < 60~\GeV$, where
the multijet contribution is most prominent, is included in the fit to
normalize the multijet control sample to the multijet contribution.  The
inclusive control sample constrains the lepton trigger and identification
efficiencies, while the pretag control sample, which includes jet
requirements, constrains a common scale factor $\kzjets$ that corrects the
$Z+$jets cross section.  The total event yields after applying all
corrections and normalization factors are shown in Table~\ref{tbl:evtall}.

\begin{table*}[!htb] 
\begin{tabular}{lccccccc}
\hline
& Data 
& Total Background
& Multijet
& $Z+$LF 
& $Z+$HF 
& Other
& $ZH$ \\
\hline \hline
inclusive  & $865254$ & $853976$ & $131905$ & $701516$ & $19074$ & $1481$ & $9.14$ \\
pretag     & $31336$ & $30634$ & $3449$ & $23234$ & $3459$ & $491$ & $6.82$ \\
ST         & $728$ & $707 \pm 130$ & $48.4$ & $161$ & $443$ & $54.1$ & $1.87 \pm 0.25$ \\
DT         & $485$ & $435 \pm 68$~ & $29.5$ & $106$ & $237$ & $61.8$ & $2.34 \pm 0.36$ \\
\hline
\end{tabular}

\caption{ 
Expected and observed event yields for all lepton channels combined after
requiring two leptons (inclusive), after also requiring two jets (pretag),
and after requiring at least one tight (ST) or two loose (DT) $b$-tags. The
total statistical and systematic uncertainties are indicated for the ``Total
Background'' and ``$ZH$'' columns of the ST and DT samples.  
The ``Other'' column includes diboson and $t\bar{t}$ event yields.  The $ZH$
sample yields are for $M_H=115~\GeV$.} \label{tbl:evtall}
\end{table*}


A multivariate analysis combines the most significant kinematic information
into a single discriminant~\cite{ancu}. Each decision tree in a random
forest (RF) \cite{tmva} is trained to separate signal from background using
a randomly selected subsample of simulated events.  In addition, a random
subset of input variables is considered for each decision in each tree.  The
RF output is a performance-weighted average of the output from each decision
tree. To exploit the kinematics of the $ZH \rightarrow \ell^+\ell^-
b\bar{b}$ process, the energies of the candidate leptons and jets are
adjusted within their experimental resolutions with a $\chi^2$ fit that
constrains $m_{\ell\ell}$ to the mass and width of the $Z$ boson, and the
$\pt$ of the $\ell^+\ell^- b\bar{b}$ system to the expected distribution for
$ZH$ events before detector resolution effects~\cite{strauss}.  The
variables selected for the RF are: the transverse momenta of the two $b$-jet
candidates and the dijet invariant mass, before and after the jet energies
are adjusted by the kinematic fit; angular differences within and between
the dijet and dilepton systems; the angle between the proton beam and the
$Z$ boson candidate in the rest frame of the $\ell^+\ell^- b\bar{b}$
system~\cite{Parke:1999}; and composite kinematic variables such as the
$\pt$ of the dijet system and the scalar sum of the transverse momenta of
the leptons and jets.  The RF outputs with all lepton channels combined are
shown separately for ST and DT events in Figs. \ref{fig:rf}(a,b).


\begin{figure*}[htbp]
\psfrag{XTITLE}[tr][tr]{{\bf RF Output}}
\centering
\begin{tabular}{ccc}
\begin{centering}
\psfrag{YTITLE}[tr][tr]{{\bf Events / 0.05}}
\psfrag{SELECTION}[tl][tl]{{\bf (a) ST}}
\psfrag{LOGO}[tl][cl]{{\bf \boldmath \mbox{D\O} 4.2 fb$^{-1}$}}
\includegraphics[height=0.16\textheight]{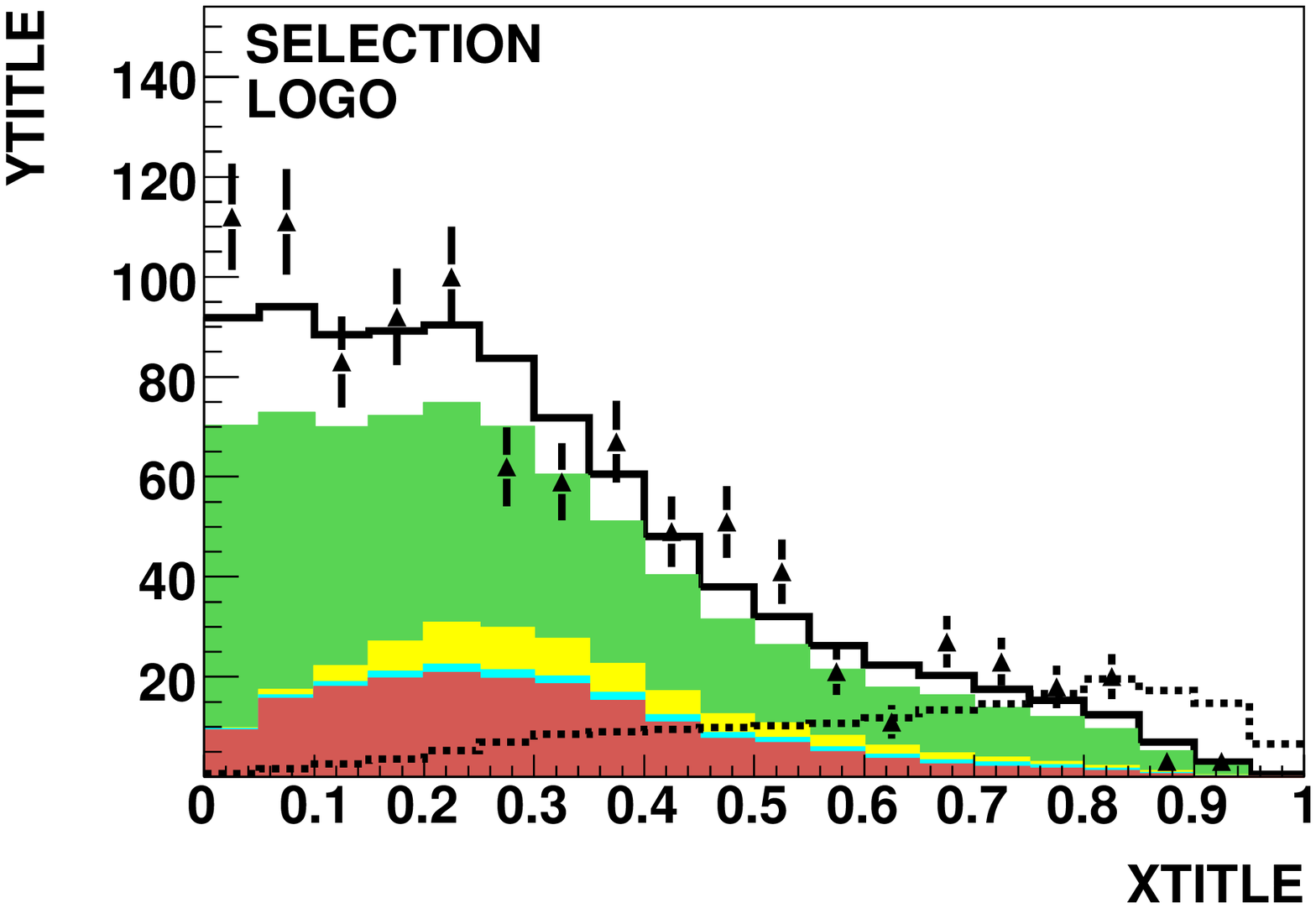} 
\end{centering} &
\begin{centering}
\psfrag{YTITLE}[tr][tr]{{\bf Events / 0.05}}
\psfrag{SELECTION}[tl][tl]{{\bf (b) DT}}
\psfrag{LOGO}[tl][cl]{{\bf \boldmath \mbox{D\O} 4.2 fb$^{-1}$}}
\psfrag{Data}[cl][cl]{{\footnotesize \bf \boldmath Data}}
\psfrag{Z+jets}[cl][cl]{{\footnotesize \bf \boldmath $Z$+LF}}
\psfrag{Z+HF}[cl][cl]{{\footnotesize \bf \boldmath $Z$+HF}}
\psfrag{Top}[cl][cl]{{\footnotesize \bf \boldmath $t\bar{t}$}}
\psfrag{Diboson}[cl][cl]{{\footnotesize \bf \boldmath Diboson}}
\psfrag{Multijet}[cl][cl]{{\footnotesize \bf \boldmath Multijet}}
\psfrag{ZH x100}[cl][cl]{{\footnotesize \bf \boldmath $ZH$x100}}
\includegraphics[height=0.16\textheight]{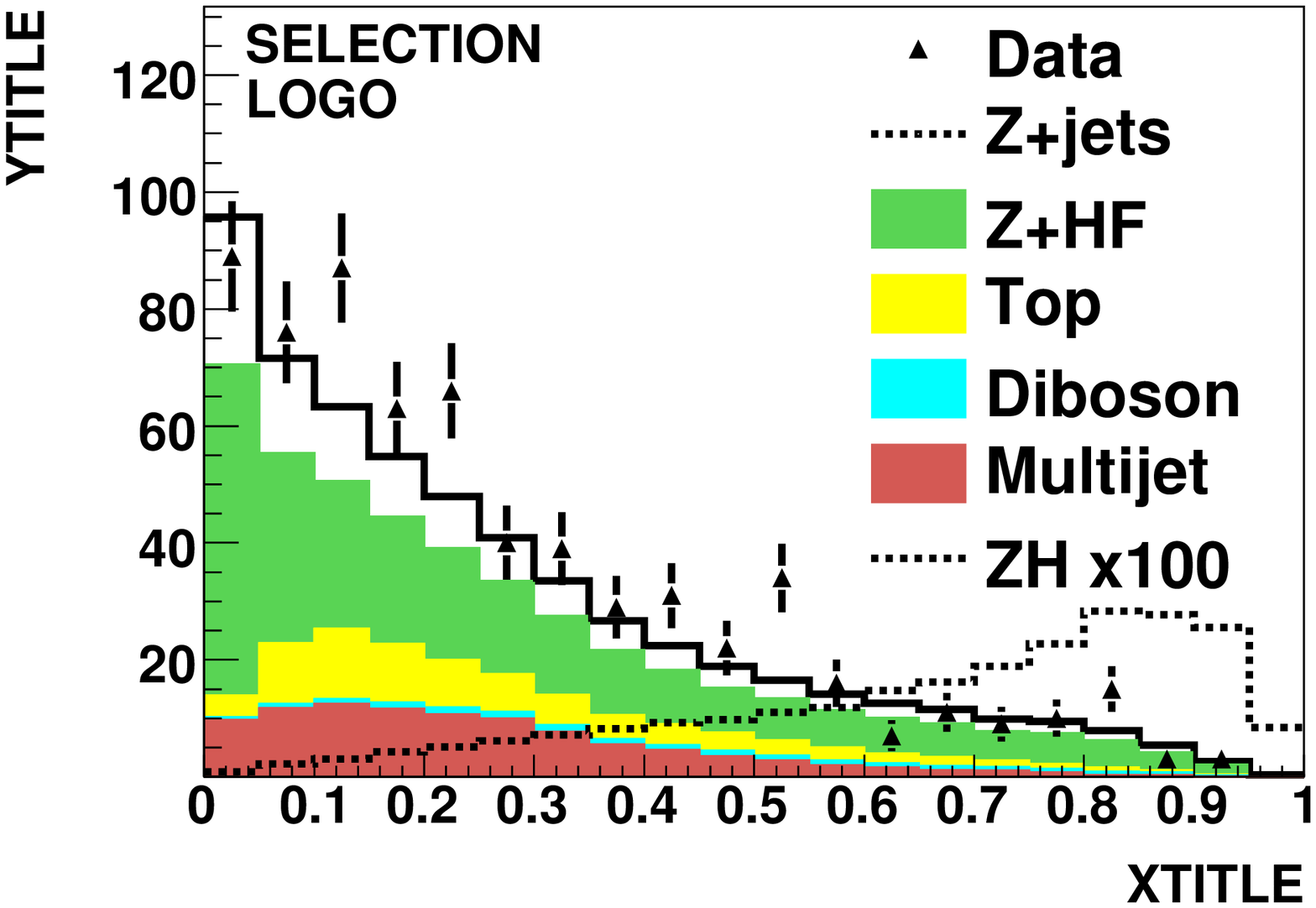} 
\end{centering} &
\begin{centering}
\psfrag{YTITLE}[tr][tr]{{\bf Events / 0.1}}
\psfrag{SELECTION}[tl][tl]{{\bf (c) ST+DT}}
\psfrag{LOGO}[br][cr]{{\bf \boldmath \mbox{D\O} 4.2 fb$^{-1}$}}
\psfrag{Data}[cl][cl]{{\footnotesize \bf \boldmath Data}}
\psfrag{ZH}[cl][cl]{{\footnotesize \bf \boldmath $ZH$x5}}
\psfrag{Pre-fit}[cl][cl]{{\footnotesize \bf \boldmath Pre-fit}}
\psfrag{Post-fit}[cl][cl]{{\footnotesize \bf \boldmath Post-fit}}
\includegraphics[height=0.16\textheight]{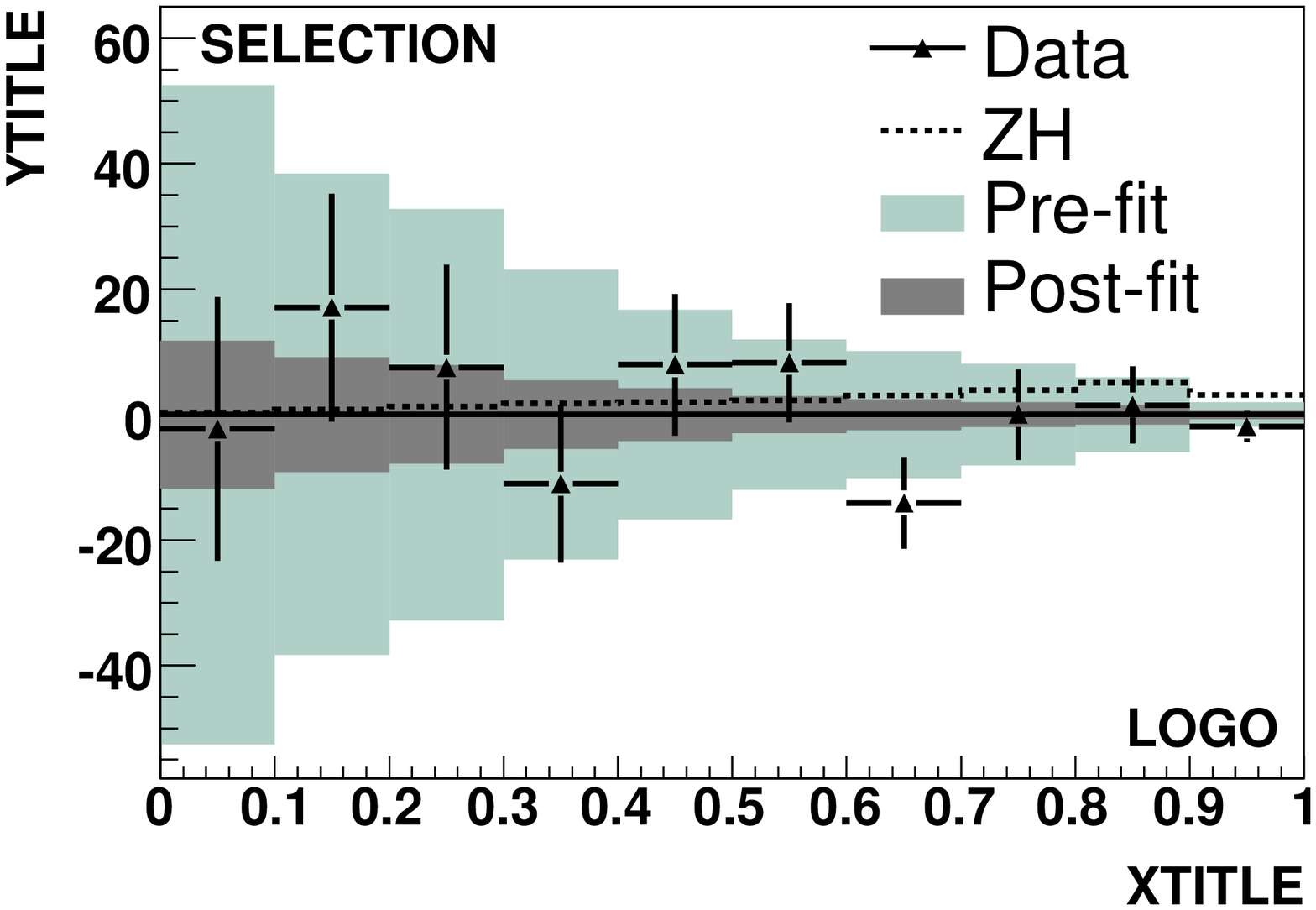}
\end{centering} \\
\end{tabular} 
\caption{\label{fig:rf} Data and background RF outputs
trained for a Higgs boson with $M_H=115~\GeV$ in (a) ST and (b) DT samples.
The (c) background-subtracted combination of ST and DT samples, with the
systematic uncertainty bands before and after the fit performed by the
limit-setting program.  }
\end{figure*}

Systematic uncertainties resulting from the background normalization are
assessed for the multijet contribution ($20$--$60\%$ depending on channel)
and for effects of lepton efficiency ($2$-$10\%$), some of which are
correlated between all lepton channels ($6\%$).  The normalization of the
$Z+$jets sample to the pretag data constrains the $Z+$jets cross section
multiplied by any jet-dependent efficiency to within the statistical
uncertainty of the pretag data ($1$--$2\%$).  Additional systematic
uncertainties ($10$--$20\%$) for possible jet-dependent efficiency effects
absorbed into $\kzjets$ are applied to the $\ttbar$, diboson and $ZH$
samples.  The normalization to the pretag data, which is dominated by
$Z$+LF, does not strongly constrain the cross sections of other processes. A
cross section uncertainty of 20\% for $Z+$HF and $6\%$--$10\%$ for other
backgrounds is determined from Ref.~\cite{mcfm}.  For the signal, the
uncertainty is 6\%~\cite{zhxsec}. The normalization to the dilepton mass
distributions reduces the impact of many of the remaining systematic
uncertainties on the background size (except those related to $b$-tagging),
but changes to the shape of the RF output distribution persist and are
accounted for. Additional sources of systematic uncertainty include: jet
energy scale, jet energy resolution, jet identification efficiency,
$b$-tagging and trigger efficiencies, PDFs, data-determined corrections to
the model for $Z+$jets, and modeling of the underlying event. The
uncertainties from the factorization and renormalization scales in the
simulation of $Z+$jets are estimated by scaling these parameters by factors
of 0.5 and 2.

No significant excess above the background expectation is observed.
Therefore, we set limits on the $ZH$ production cross section with a
modified frequentist (CLs) method that uses a negative log likelihood ratio
(LLR) of the signal-plus-background (S+B) hypothesis to the background-only
(B) hypothesis \cite{cls}.  The RF output distributions and corresponding systematic uncertainties of
the ST and DT samples from each leptonic channel and from two distinct data
taking periods are analyzed separately by the limit setting program to take
advantage of the sensitivity in the more discriminating channels.  
To minimize the impact of the systematic uncertainties, the likelihood of
the B and S+B hypotheses are each maximized by independent fits that vary
nuisance parameters used to model the systematic effects \cite{wade}.  The
correlations among systematic uncertainties are maintained across channels,
as well as backgrounds and signal.  The background-subtracted RF
distribution, combined for all channels, with systematic uncertainty bands
both before and after the fitting procedure, is shown in Fig.~\ref{fig:rf}c.

\begin{figure}[t]
\psfrag{YTITLE}[tr][tr]{{\bf \boldmath LLR}} \psfrag{XTITLE}[tr][tr]{{\bf
\boldmath $M_H~[\GeV]$}} \psfrag{LOGO}[tr][tr]{{\bf \boldmath \mbox{D\O} 4.2
fb$^{-1}$}} \psfrag{CHANNEL}[tr][cr]{{\bf \boldmath $ZH \rightarrow llbb$}}
\psfrag{LB2}[lc][lc]{{\bf \boldmath B$\pm$2 s.d.}} \psfrag{LB1}[lc][lc]{{\bf
\boldmath B$\pm$1 s.d.}} \psfrag{LBE}[lc][lc]{{\bf \boldmath B}}
\psfrag{LBSE}[lc][lc]{{\bf \boldmath S$+$B}} \psfrag{LOBS}[lc][lc]{{\bf
\boldmath Observed}}
\includegraphics[height=0.22\textheight]{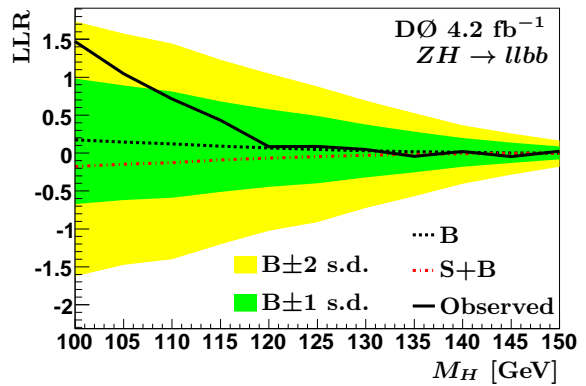}
\caption{ Observed LLR as a function of Higgs boson mass.  Also shown are 
the expected LLRs for the B and S+B hypotheses, together with the one and
two standard deviation (s.d.) bands of the background-only expectation.} \label{fig:results}
\end{figure}

\begin{table*}[htdp]
\begin{center}
\begin{tabular}{lccccccccccc}
\hline
 $M_H~(\GeV)$ & 100 & 105 & 110 & 115 & 120 & 125 & 130 & 135 & 140 & 145 & 150 \\
\hline\hline
Expected/SM:      & 5.1 & 5.6 & 6.2 & 7.1 & 8.4 & 10.0 & 12.7 & 16.8 & 23.6 & 34.0 & 53.6 \\
Observed/SM:      & 3.0 & 3.8 & 4.6 & 5.9 & 7.9 & 9.2 & 12.1 & 17.3 & 22.3 & 39.4 & 49.3 \\
Observed (fb):    & 41 & 44 & 44 & 47 & 50 & 45 & 45 & 46 & 41 & 47 & 36 \\
\hline
\end{tabular}
\caption{The expected and observed 95\% CL upper limits on the SM Higgs
boson production cross section for $ZH \rightarrow \ell^+\ell^- b\bar{b}$,
expressed as a ratio to the SM cross section.  The corresponding observed
limits on the $ZH$ production cross section multiplied by the branching ratio of $H
\rightarrow b\bar{b}$ are also reported (in fb).\label{tbl:limits}}
\end{center}
\end{table*}

Figure~\ref{fig:results} shows the observed LLR as a function of Higgs boson
mass.  Also shown are the expected (median) LLRs for the B and S+B
hypotheses, together with the one and two standard deviation bands of the
background-only expectation. A signal-like excess would result in a negative
value of observed LLR.  The data are consistent with either hypothesis for
the entire mass range $100 < M_H < 150~\GeV$.  The 95\% CL upper limit on
the cross section times branching ratio, expressed as a ratio to the SM
prediction, for each $M_H$ is presented in Table~\ref{tbl:limits}.  At
$M_H=115~\GeV$, the observed (expected) limit on this ratio is 5.9 (7.1).
Compared to the previous best expected limit in this channel \cite{cdfzh},
this represents a $40\%$ improvement.

Supplementary material detailing the pretag control sample, the effect of
the kinematic fit, and additional cross section limits and LLR distributions
from individual lepton channels is available at \cite{appendix}.

%
We thank the staffs at Fermilab and collaborating institutions,
and acknowledge support from the
DOE and NSF (USA);
CEA and CNRS/IN2P3 (France);
FASI, Rosatom and RFBR (Russia);
CNPq, FAPERJ, FAPESP and FUNDUNESP (Brazil);
DAE and DST (India);
Colciencias (Colombia);
CONACyT (Mexico);
KRF and KOSEF (Korea);
CONICET and UBACyT (Argentina);
FOM (The Netherlands);
STFC and the Royal Society (United Kingdom);
MSMT and GACR (Czech Republic);
CRC Program and NSERC (Canada);
BMBF and DFG (Germany);
SFI (Ireland);
The Swedish Research Council (Sweden);
and
CAS and CNSF (China).
%


\begin{table*}
\begin{center}
{\Large{\bf Supplementary Material}}
\end{center}
\end{table*}

\begin{figure*}[htbp]
\centering
\begin{tabular}{cc}
\begin{centering}
\psfrag{LOGO}[tl][cl]{{\bf \boldmath \mbox{D\O} 4.2 fb$^{-1}$}}
\psfrag{YTITLE}[tr][tr]{{\bf \boldmath Events / 1.5~$\GeV$}}
\psfrag{XTITLE}[tr][tr]{{\bf \boldmath Dilepton Mass [$\GeV$]}}
\psfrag{SELECTION}[tl][tl]{{\bf \boldmath (a) Pretag}}
\includegraphics[height=0.20\textheight]{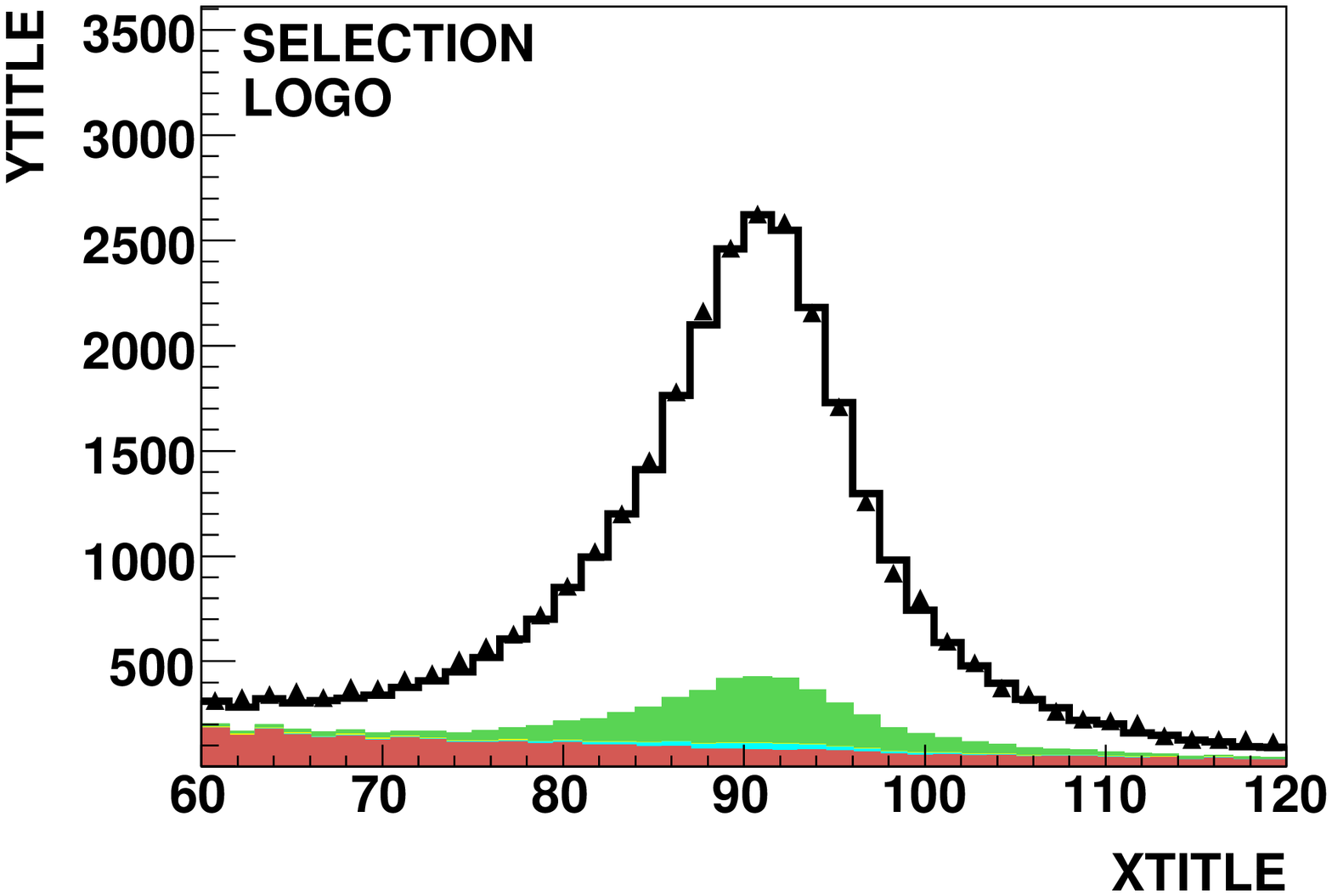}
\end{centering} &
\begin{centering}
\psfrag{LOGO}[tl][cl]{{\bf \boldmath \mbox{D\O} 4.2 fb$^{-1}$}}
\psfrag{YTITLE}[tr][tr]{{\bf \boldmath Events / 5~$\GeV$}}
\psfrag{SELECTION}[tl][tl]{{\bf \boldmath (b) Pretag}}
\psfrag{XTITLE}[tr][tr]{{\bf \boldmath Dijet Mass [$\GeV$]}}
\psfrag{Data}[cl][cl]{{\footnotesize \bf \boldmath Data}}
\psfrag{Z+LF}[cl][cl]{{\footnotesize \bf \boldmath $Z$+LF}}
\psfrag{Z+HF}[cl][cl]{{\footnotesize \bf \boldmath $Z$+HF}}
\psfrag{Top}[cl][cl]{{\footnotesize \bf \boldmath $t\bar{t}$}}
\psfrag{Diboson}[cl][cl]{{\footnotesize \bf \boldmath Diboson}}
\psfrag{Multijet}[cl][cl]{{\footnotesize \bf \boldmath Multijet}}
\includegraphics[height=0.20\textheight]{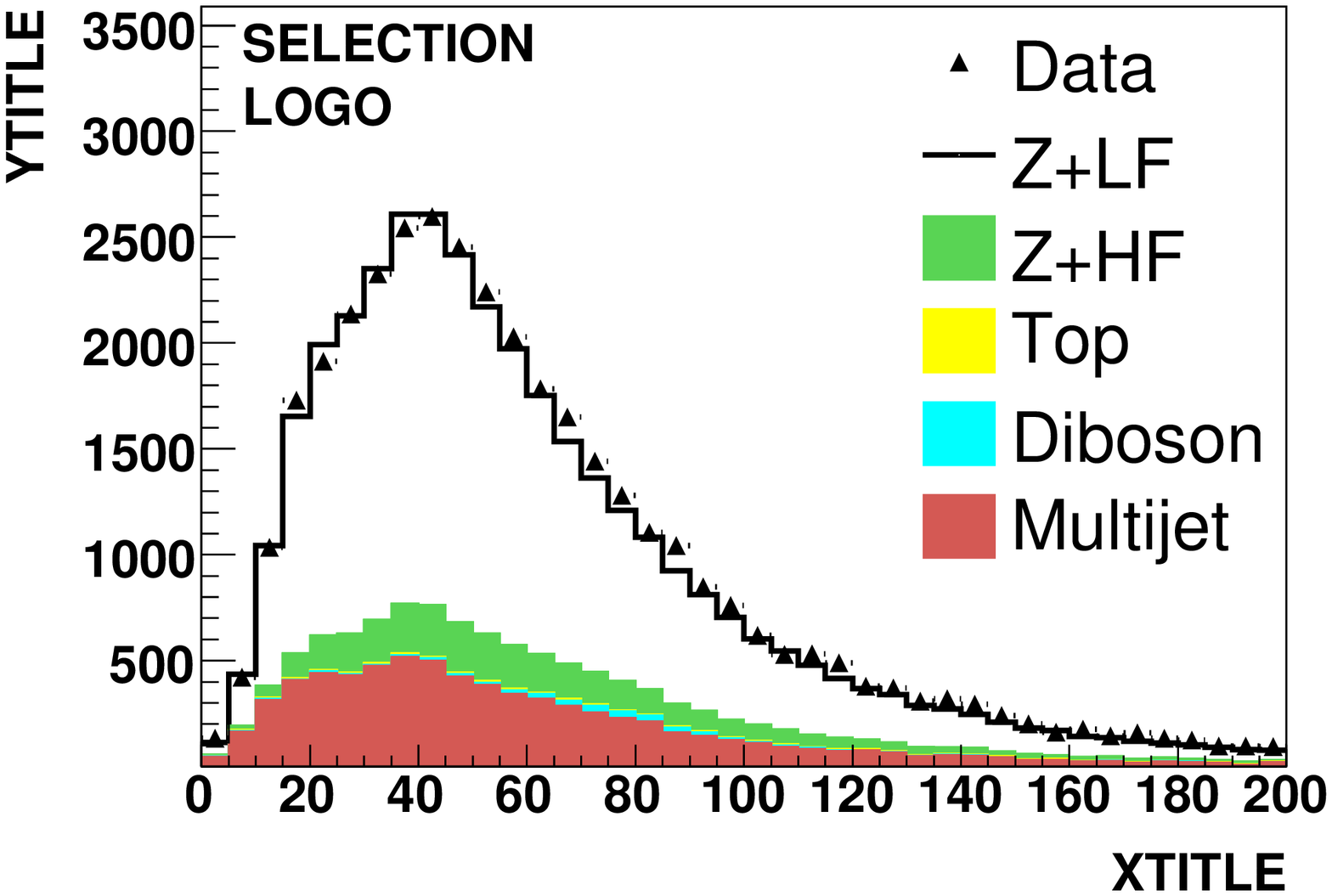}
\end{centering} \\
\begin{centering}
\psfrag{LOGO}[tl][cl]{{\bf \boldmath \mbox{D\O} 4.2 fb$^{-1}$}}
\psfrag{YTITLE}[tr][tr]{{\bf \boldmath Events / 0.025}}
\psfrag{XTITLE}[tr][tr]{{\bf \boldmath RF Output}}
\psfrag{SELECTION}[tl][tl]{{\bf \boldmath (c) ST-trained, Pretag}}
\includegraphics[height=0.20\textheight]{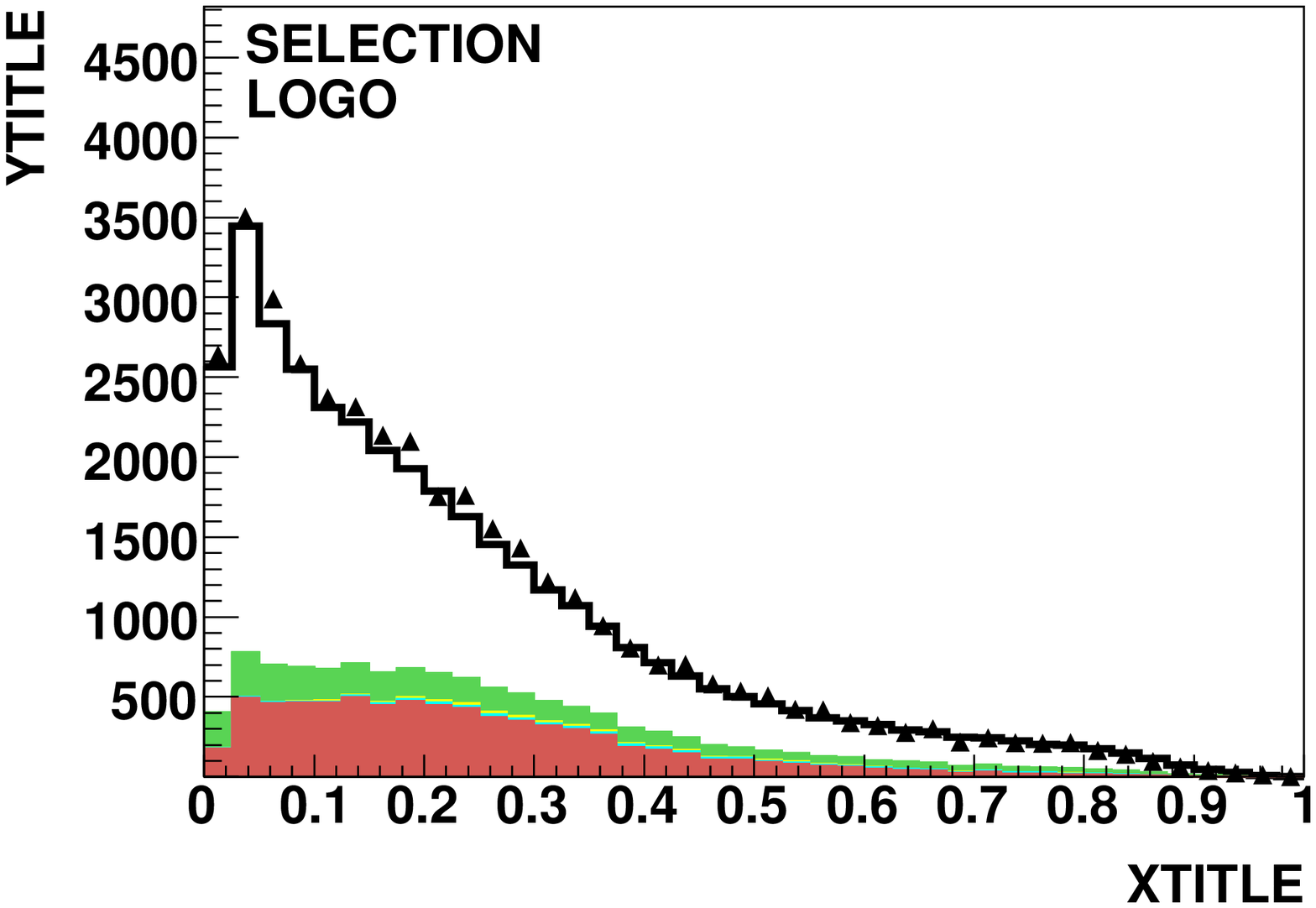}
\end{centering} &
\begin{centering}
\psfrag{LOGO}[tl][cl]{{\bf \boldmath \mbox{D\O} 4.2 fb$^{-1}$}}
\psfrag{YTITLE}[tr][tr]{{\bf \boldmath Events / 0.025}}
\psfrag{XTITLE}[tr][tr]{{\bf \boldmath RF Output}}
\psfrag{SELECTION}[tl][tl]{{\bf \boldmath (d) DT-trained, Pretag}}
\includegraphics[height=0.20\textheight]{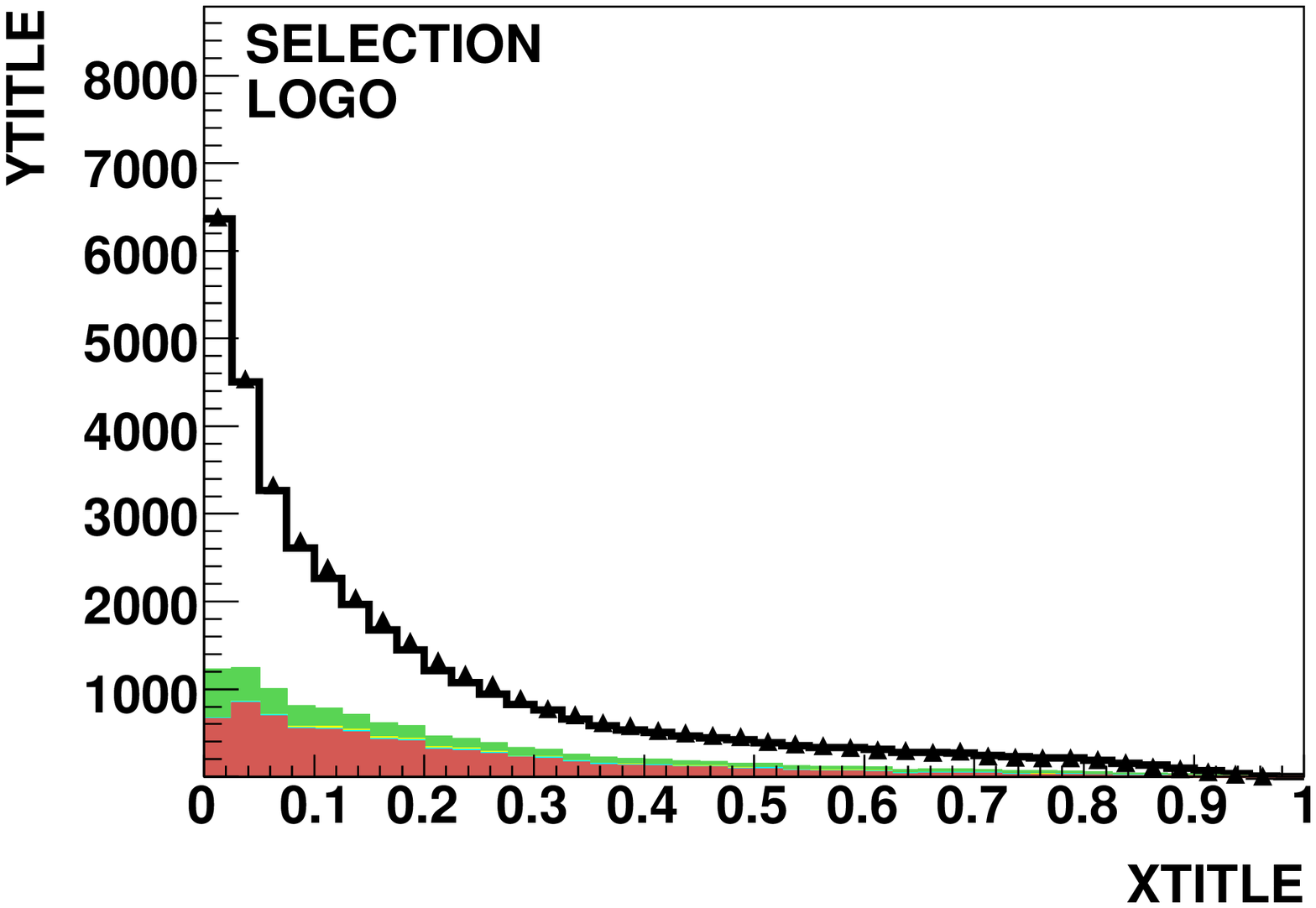}
\end{centering} \\
\begin{centering}
\psfrag{LOGO}[tr][cr]{{\bf \boldmath \mbox{D\O} 4.2 fb$^{-1}$}}
\psfrag{YTITLE}[tr][tr]{{\bf \boldmath Events / 0.025}}
\psfrag{XTITLE}[tr][tr]{{\bf \boldmath RF Output}}
\psfrag{SELECTION}[tr][tr]{{\bf \boldmath (e) ST-trained, Pretag}}
\includegraphics[height=0.20\textheight]{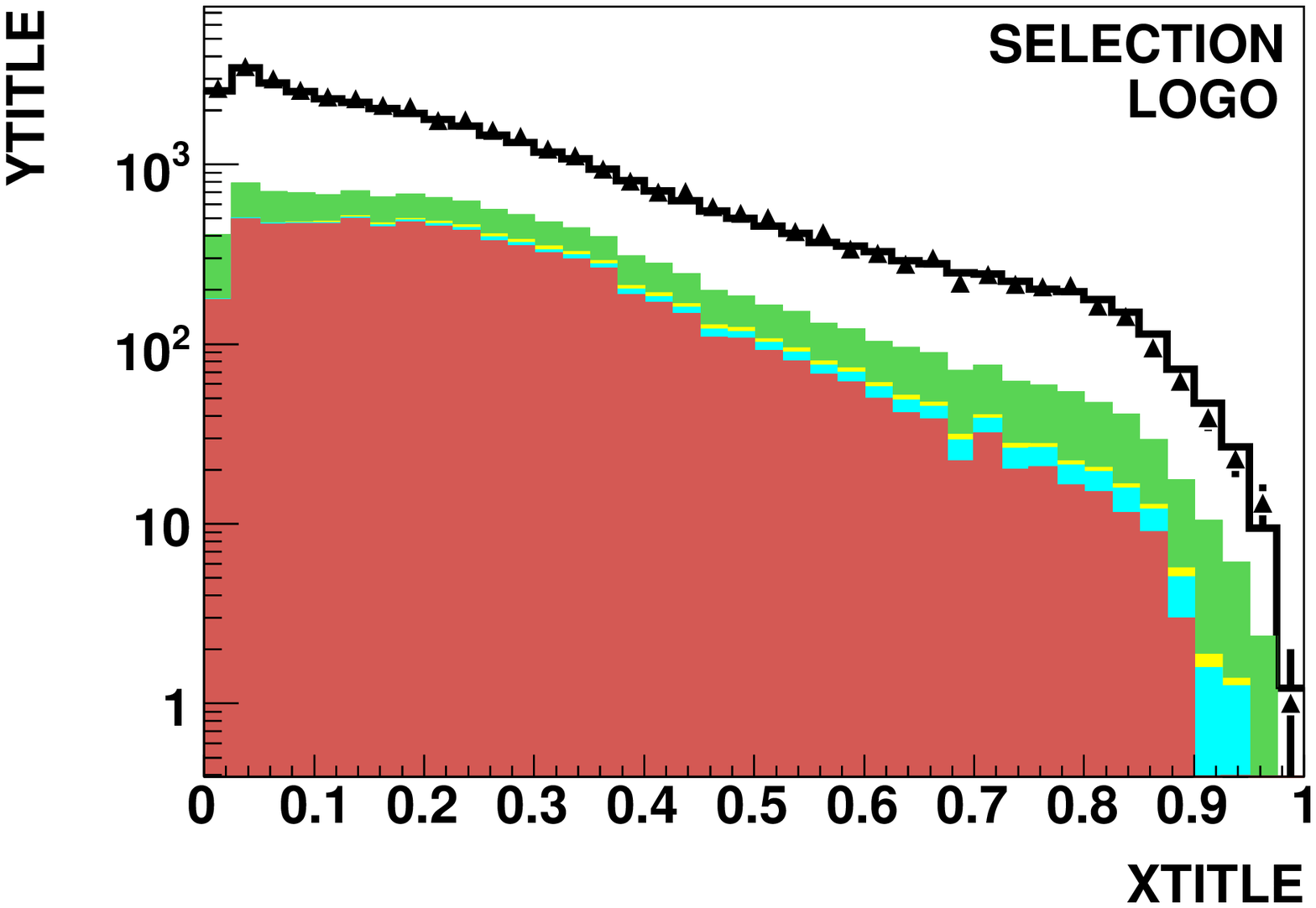}
\end{centering} &
\begin{centering}
\psfrag{LOGO}[tr][cr]{{\bf \boldmath \mbox{D\O} 4.2 fb$^{-1}$}}
\psfrag{YTITLE}[tr][tr]{{\bf \boldmath Events / 0.025}}
\psfrag{XTITLE}[tr][tr]{{\bf \boldmath RF Output}}
\psfrag{SELECTION}[tr][tr]{{\bf \boldmath (f) DT-trained, Pretag}}
\includegraphics[height=0.20\textheight]{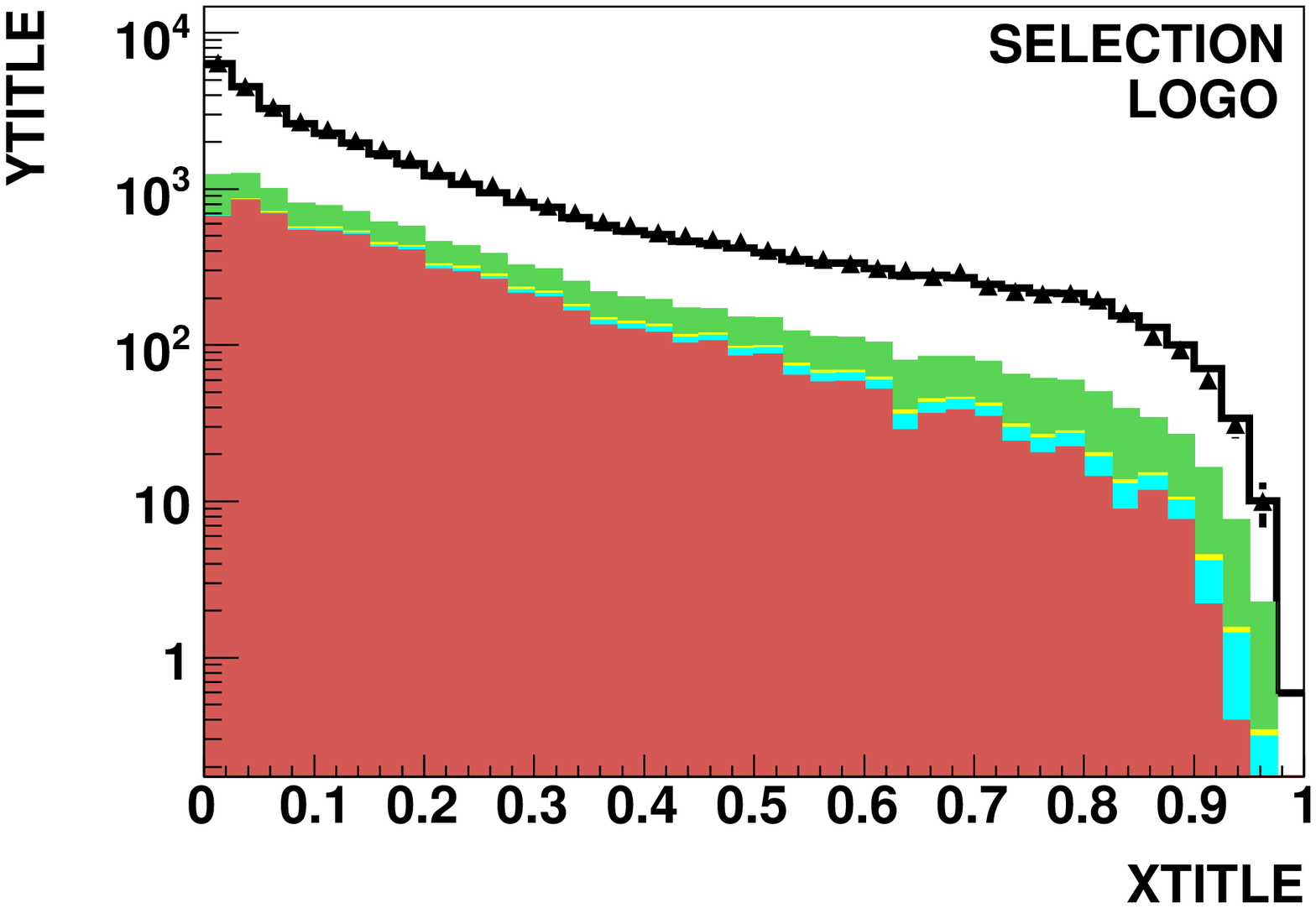}
\end{centering} \\
\end{tabular}
\caption{ Pretag distributions for (a) the dilepton invariant mass, (b) the dijet
invariant mass after the kinematic fit, (c) the RF discriminant trained for
ST events, (d) the RF discriminant trained for DT events, for all lepton channels combined.
(e) and (f) reproduce (c) and (d) using a logarithmic scale. } \label{fig:pretag_dists}
\end{figure*}

\begin{figure*}[htbp]
\psfrag{YTITLE}[tr][tr]{{\bf \boldmath Events / 10~$\GeV$}}
\psfrag{LOGO}[tl][cl]{{\bf \boldmath \mbox{D\O} 4.2 fb$^{-1}$}}
\centering
\begin{tabular}{cc}
\begin{centering}
\psfrag{SELECTION}[tl][tl]{{\bf \boldmath (a) ST}}
\psfrag{XTITLE}[tr][tr]{{\bf \boldmath Dijet Mass [$\GeV$]}}
\includegraphics[height=0.20\textheight]{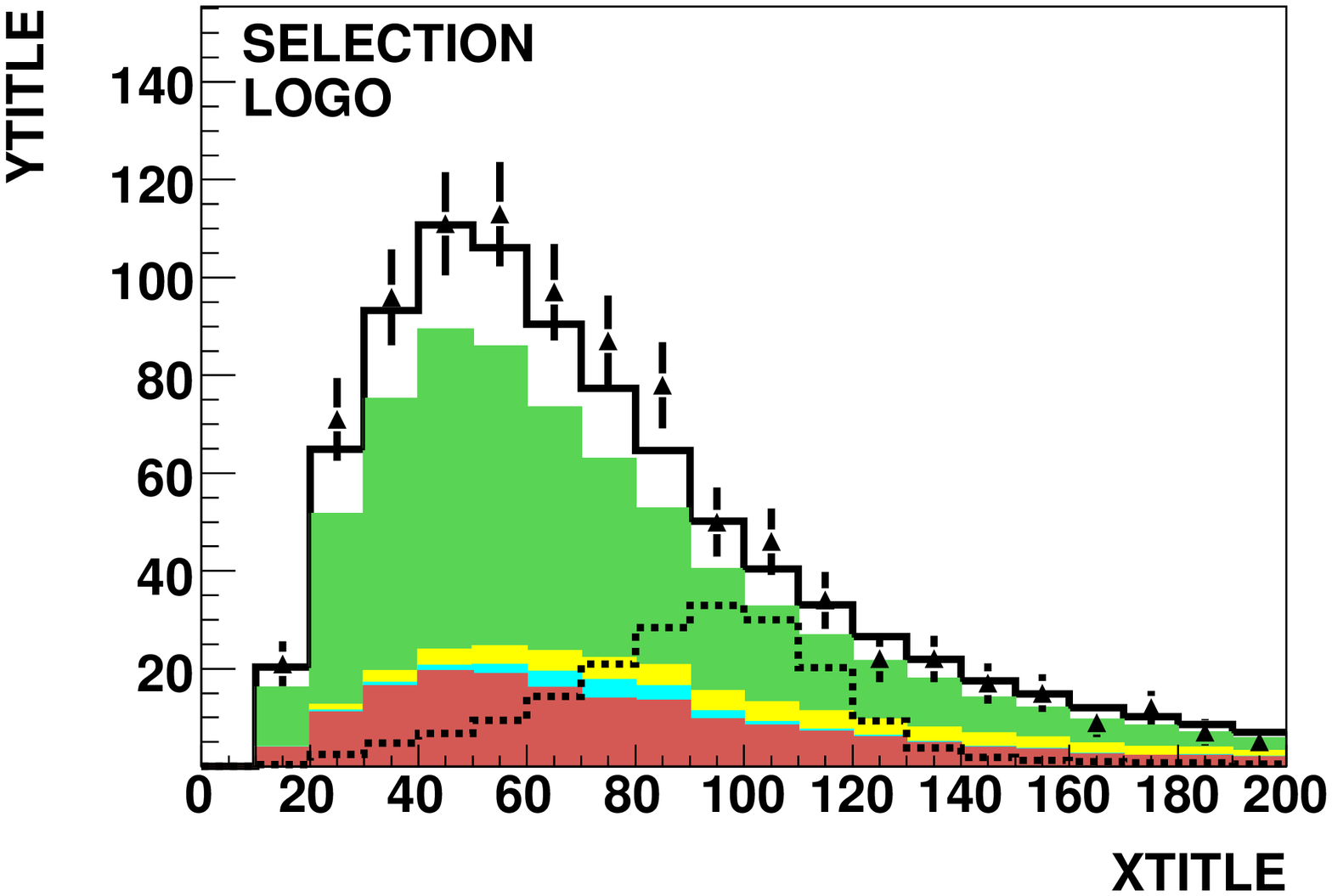}
\end{centering} &
\begin{centering}
\psfrag{SELECTION}[tl][tl]{{\bf \boldmath (b) DT}}
\psfrag{XTITLE}[tr][tr]{{\bf \boldmath Dijet Mass [$\GeV$]}}
\psfrag{Data}[cl][cl]{{\footnotesize \bf \boldmath Data}}
\psfrag{Z+jets}[cl][cl]{{\footnotesize \bf \boldmath $Z$+LF}}
\psfrag{Z+HF}[cl][cl]{{\footnotesize \bf \boldmath $Z$+HF}}
\psfrag{Top}[cl][cl]{{\footnotesize \bf \boldmath $t\bar{t}$}}
\psfrag{Diboson}[cl][cl]{{\footnotesize \bf \boldmath Diboson}}
\psfrag{Multijet}[cl][cl]{{\footnotesize \bf \boldmath Multijet}}
\psfrag{ZH x100}[cl][cl]{{\footnotesize \bf \boldmath $ZH$x100}}
\includegraphics[height=0.20\textheight]{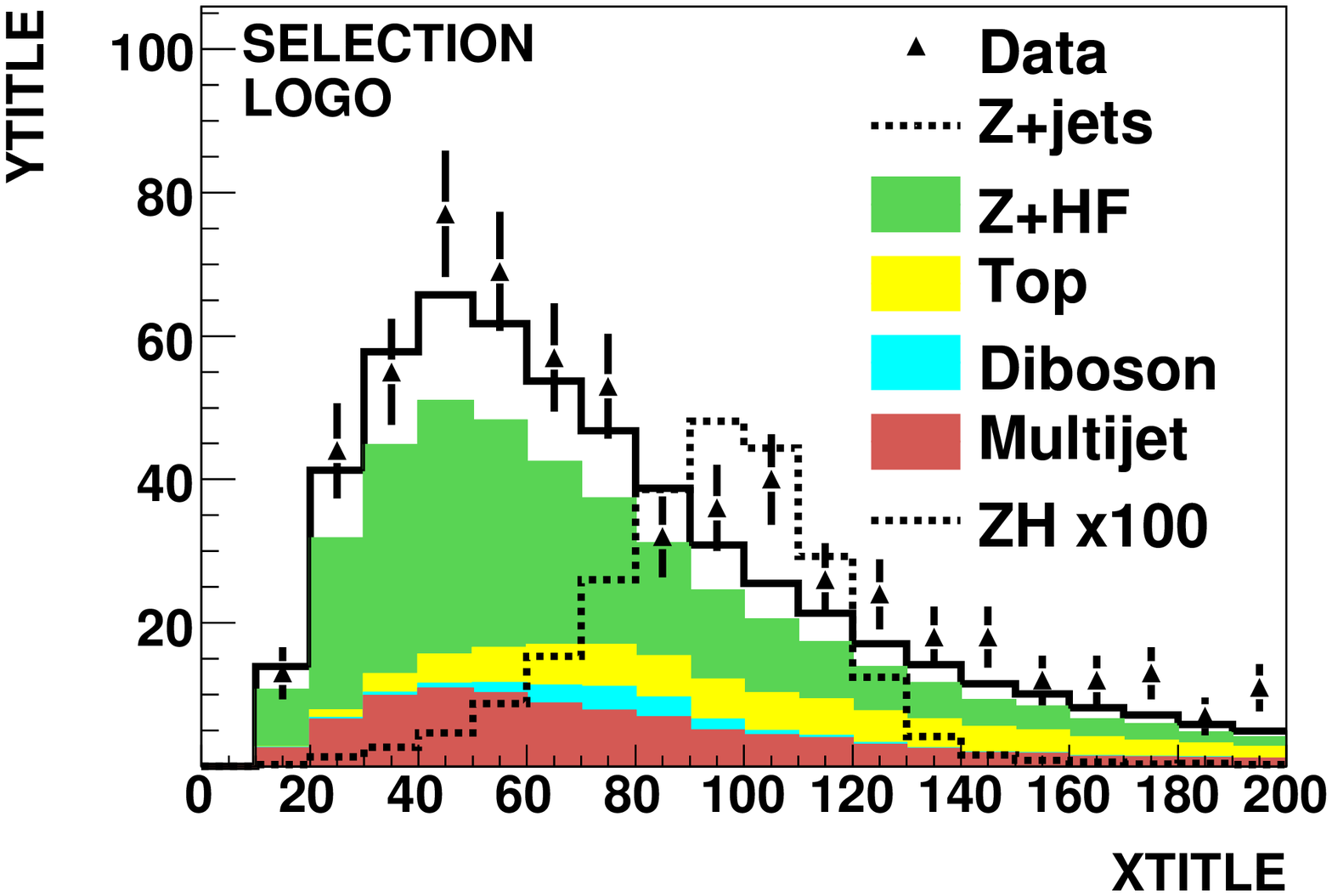}
\end{centering} \\
\begin{centering}
\psfrag{SELECTION}[tl][tl]{{\bf \boldmath (c) ST, Kinematic fit}}
\psfrag{XTITLE}[tr][tr]{{\bf \boldmath Dijet Mass [$\GeV$]}}
\includegraphics[height=0.20\textheight]{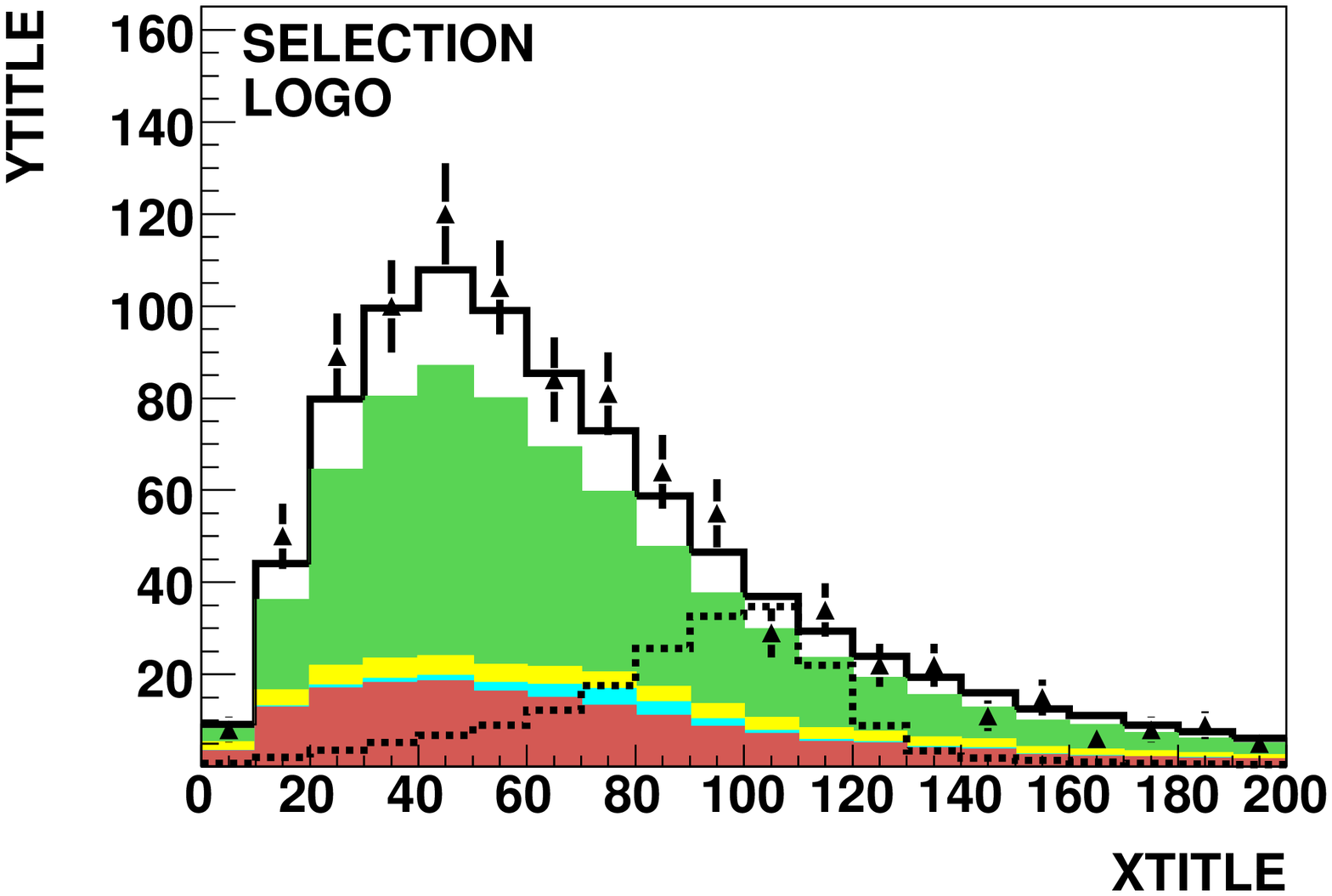}
\end{centering} &
\begin{centering}
\psfrag{SELECTION}[tl][tl]{{\bf \boldmath (d) DT, Kinematic fit}}
\psfrag{XTITLE}[tr][tr]{{\bf \boldmath Dijet Mass [$\GeV$]}}
\includegraphics[height=0.20\textheight]{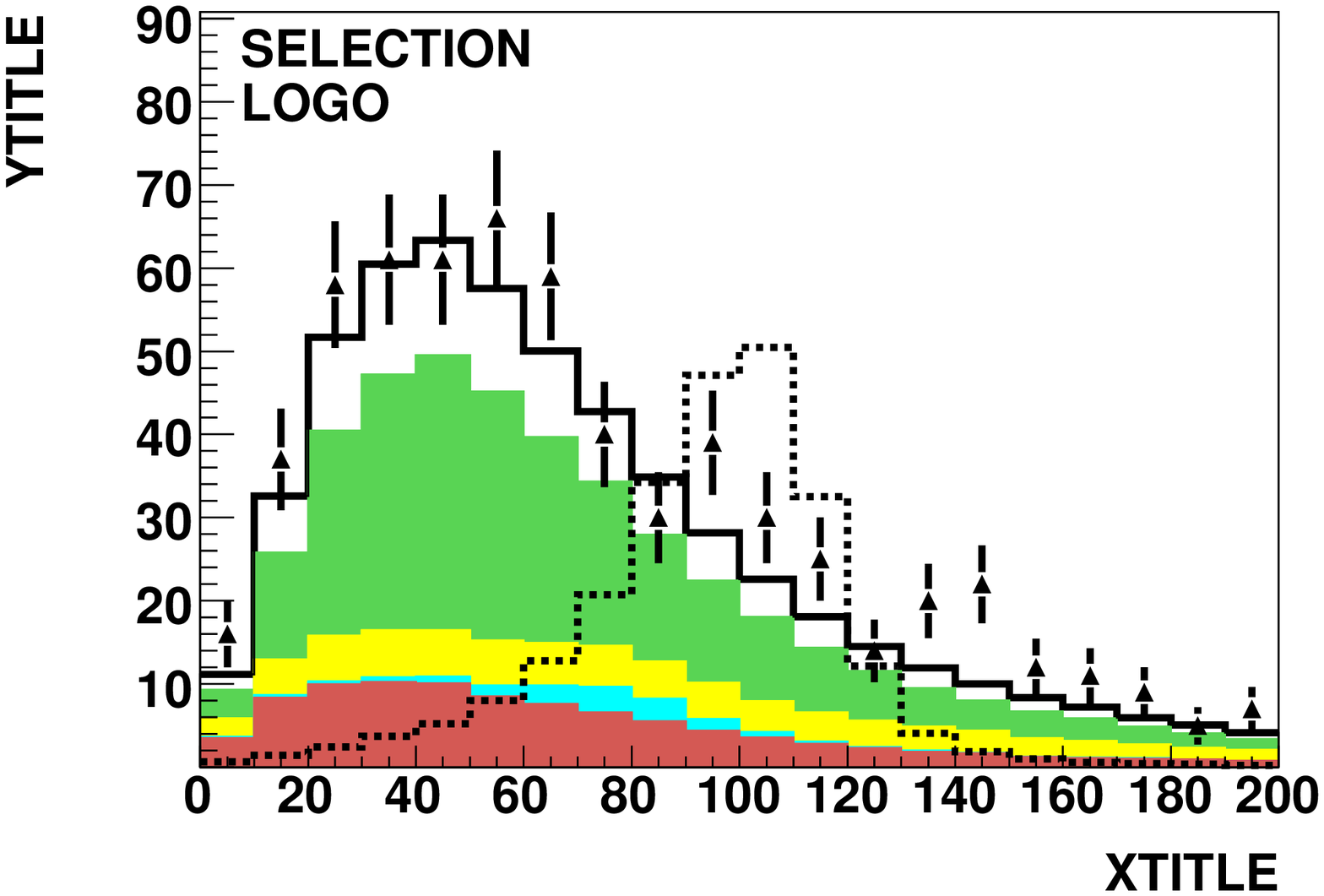}
\end{centering} \\
\end{tabular}
\caption{Dijet invariant mass distributions before the kinematic fit in (a)
ST events, and (b) DT events; and calculated from jet energies as adjusted
by the kinematic fit in (c) ST events and (d) DT events combined for all
lepton channels.  The $ZH$ signal shown is for $M_H=115~\GeV$.} \label{fig:mbb_dists}
\end{figure*}

\begin{figure*}[htbp]
\psfrag{YTITLE}[tr][tr]{{\bf \boldmath Limit / SM}}
\psfrag{XTITLE}[tr][tr]{{\bf \boldmath $M_H$~[$\GeV$]}}
\psfrag{CHANNEL}[tl][cl]{{\bf \boldmath \mbox{D\O} 4.2 fb$^{-1}$}}
\psfrag{Observed Limit}[tl][tl]{{\bf \boldmath Observed}}
\psfrag{Expected Limit}[tl][cl]{{\bf \boldmath Expected}}
\begin{tabular}{cc}
\begin{centering}
\psfrag{LOGO}[tl][tl]{{\bf \boldmath (a) $ZH \rightarrow \ell \ell bb$ }}
\includegraphics[height=0.22\textheight]{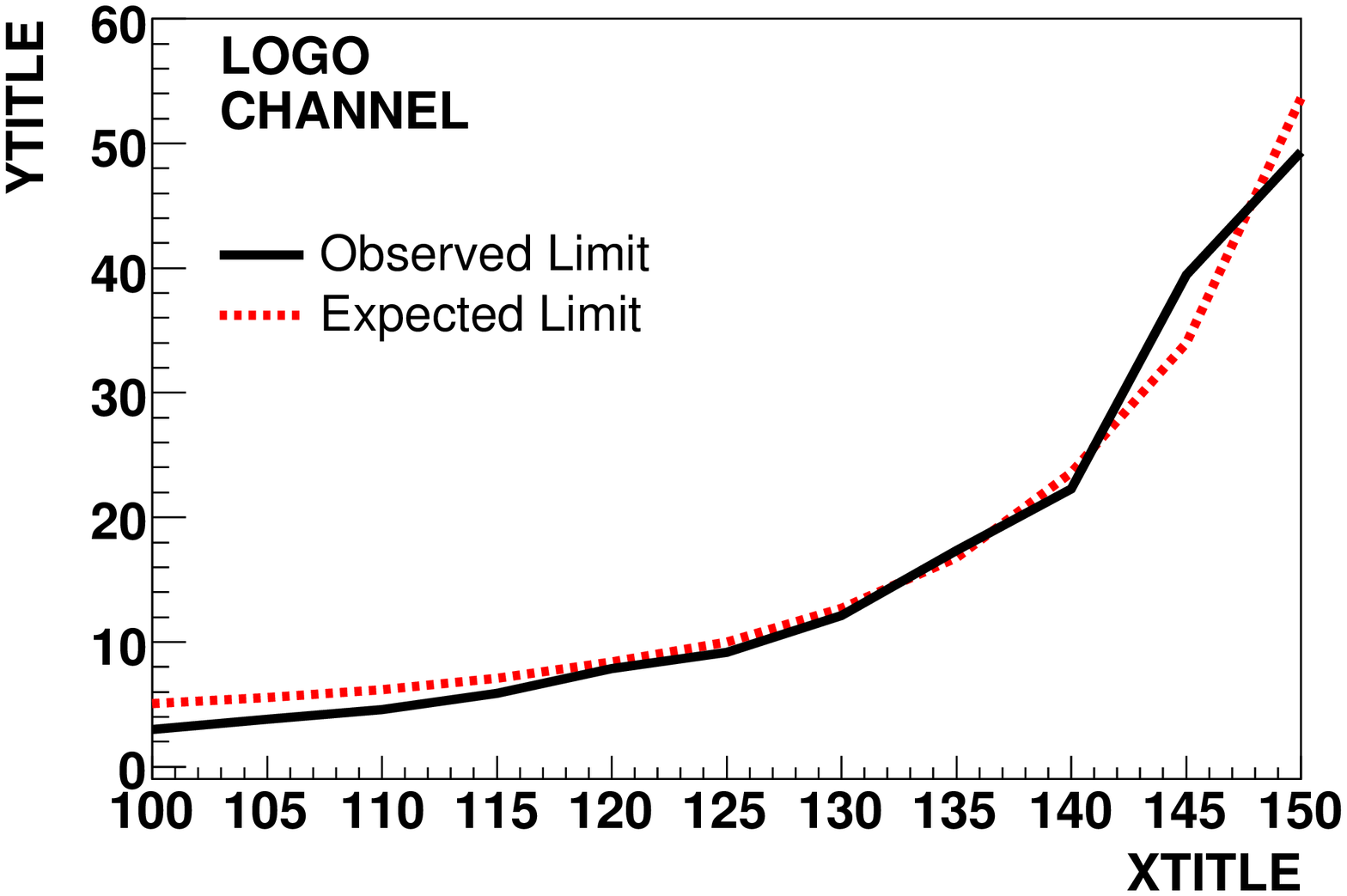}
\end{centering} &
\begin{centering}
\psfrag{LOGO}[tl][tl]{{\bf \boldmath (b) $ZH \rightarrow \ell \ell bb$}}
\includegraphics[height=0.22\textheight]{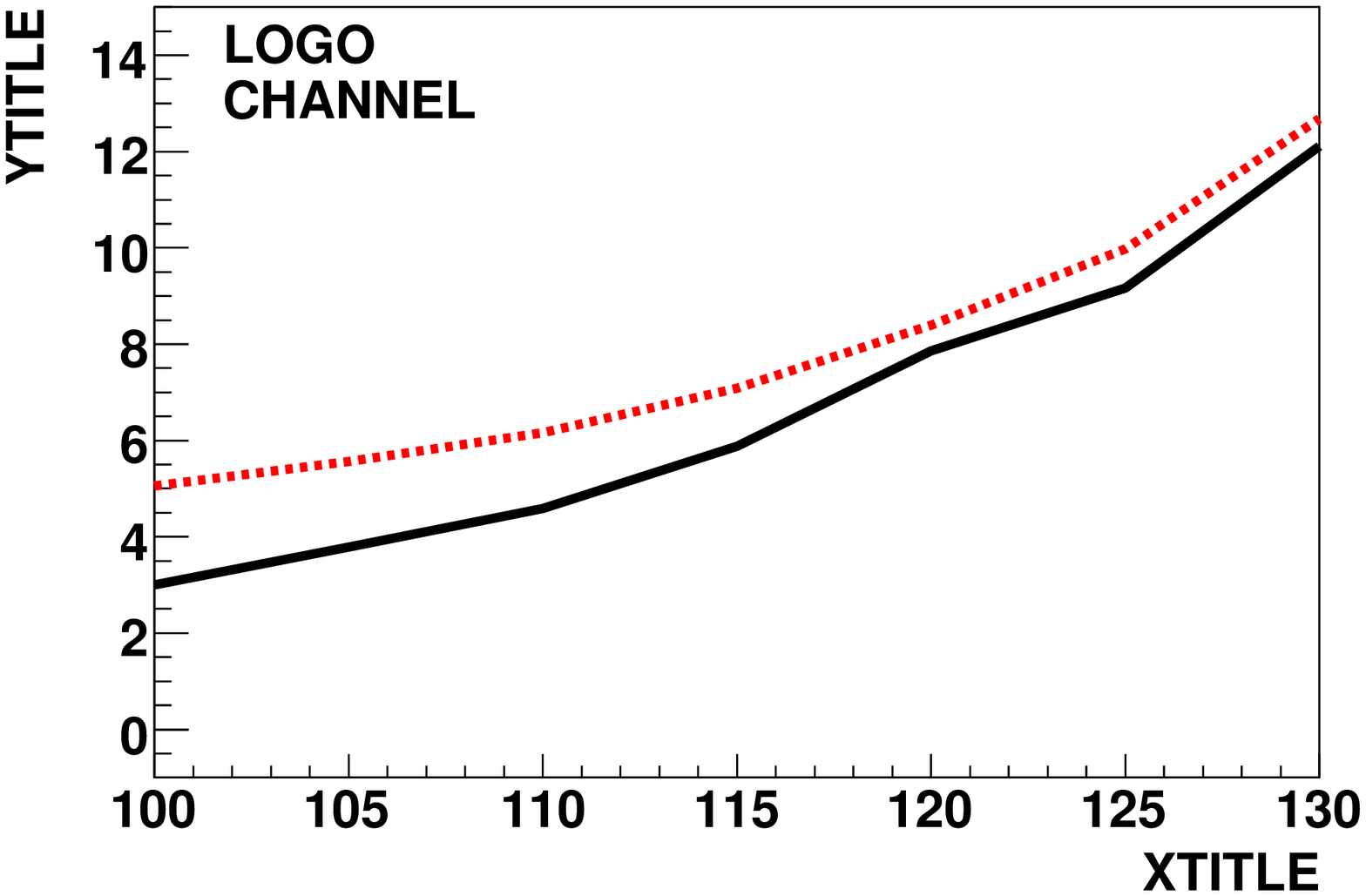}
\end{centering} \\
\end{tabular}
\caption{ The expected and observed 95\% C.L. cross section limit divided by the SM
Higgs boson production cross section as a function of $M_H$ (a) for $M_H\leq
150~\GeV$ and (b) for $M_H\leq 130~\GeV$. Limits are for the combination of the
DT and ST samples in all lepton channels.} \label{fig:lim}
\end{figure*}

\begin{figure*}[htbp]
\psfrag{YTITLE}[tr][tr]{{\bf \boldmath LLR}}
\psfrag{XTITLE}[tr][tr]{{\bf \boldmath $M_H~[\GeV]$}}
\psfrag{CHANNEL}[tr][br]{{\bf \boldmath \mbox{D\O} 4.2 fb$^{-1}$}}
\psfrag{LB2}[lc][lc]{{\bf \boldmath B$\pm$2 s.d.}}
\psfrag{LB1}[lc][lc]{{\bf \boldmath B$\pm$1 s.d.}}
\psfrag{LBE}[lc][lc]{{\bf \boldmath B}}
\psfrag{LBSE}[lc][lc]{{\bf \boldmath B+S}}
\psfrag{LOBS}[lc][lc]{{\bf \boldmath Observed}}
\begin{tabular}{cc}
\begin{centering}
\psfrag{LOGO}[tr][tr]{{\bf \boldmath (a) $ZH \rightarrow eebb$}}
\includegraphics[height=0.20\textheight]{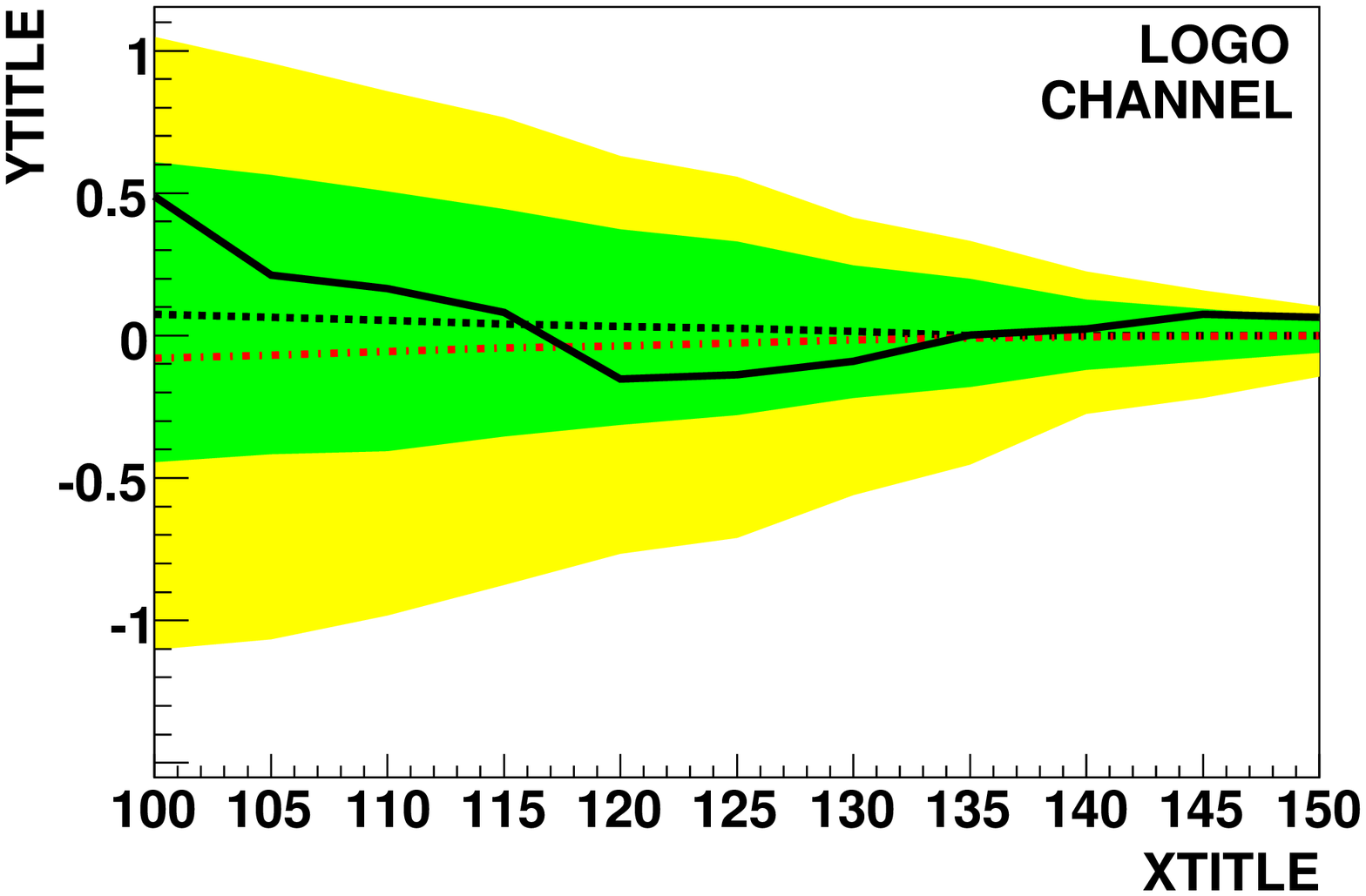}
\end{centering} &
\begin{centering}
\psfrag{LOGO}[tr][tr]{{\bf \boldmath (b) $ZH \rightarrow \mu\mu bb$}}
\includegraphics[height=0.20\textheight]{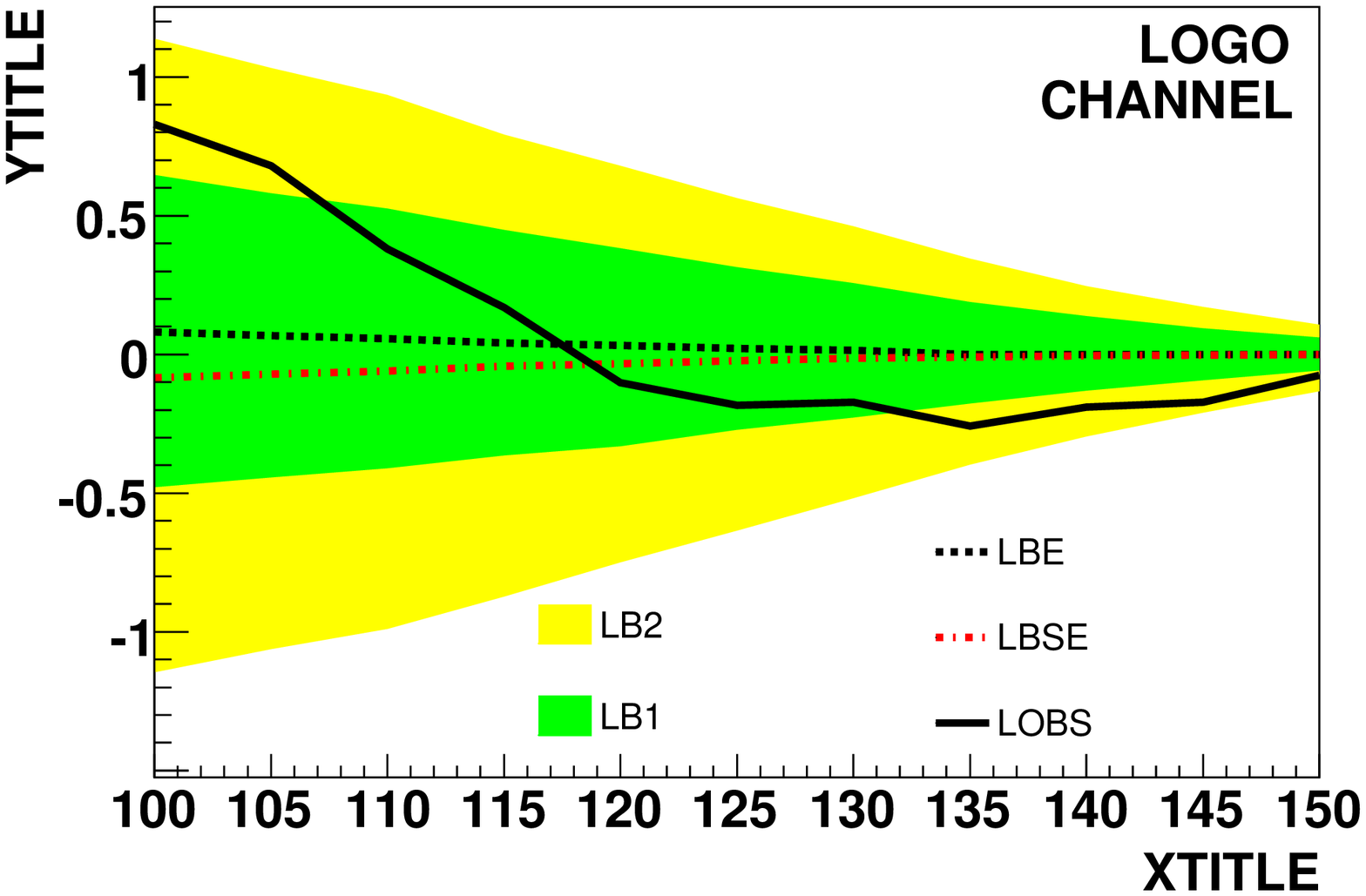}
\end{centering} \\
\begin{centering}
\psfrag{LOGO}[tr][tr]{{\bf \boldmath (c) $ZH \rightarrow \eeicr bb$}}
\includegraphics[height=0.20\textheight]{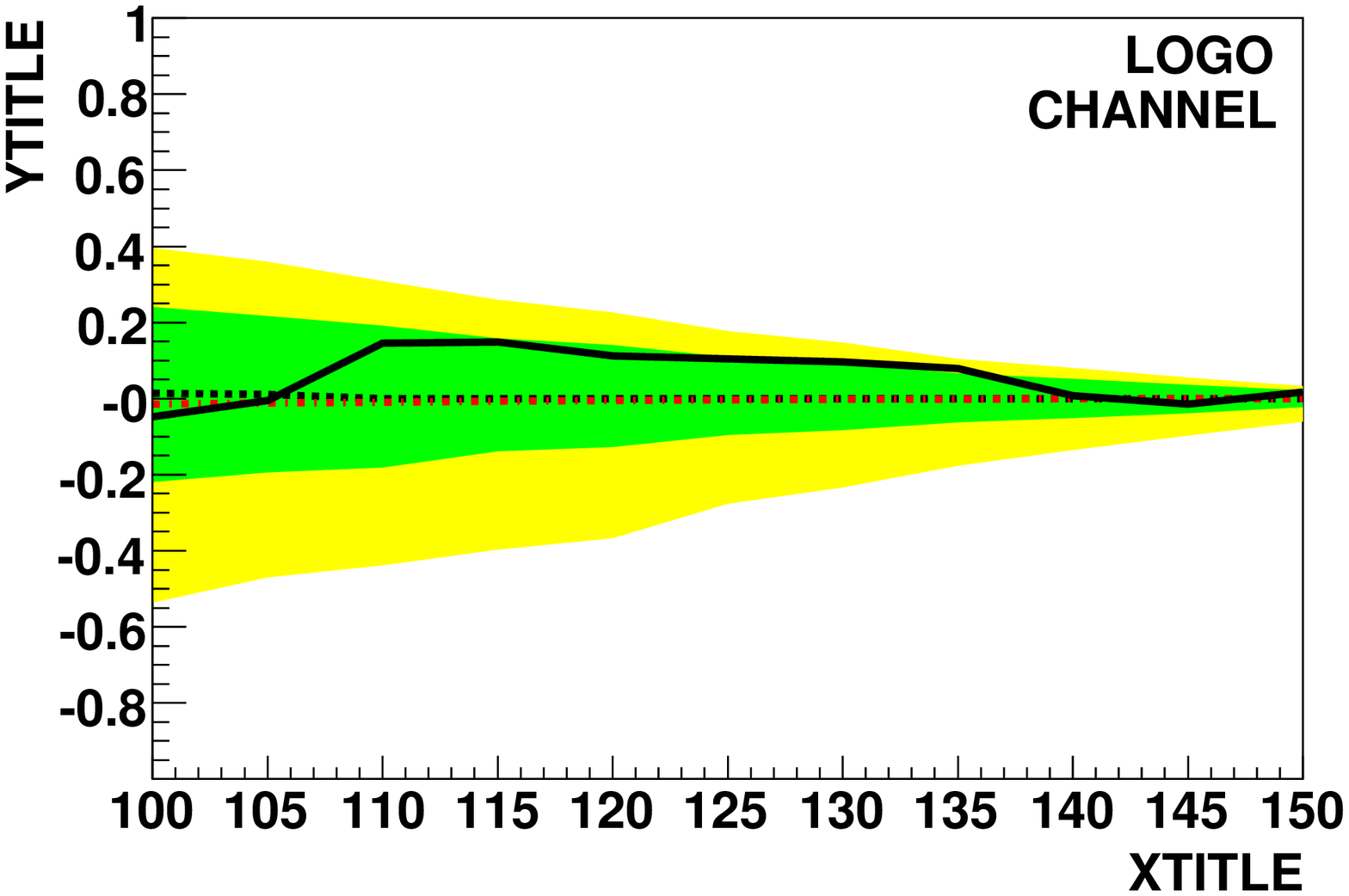}
\end{centering} &
\begin{centering}
\psfrag{LOGO}[tr][tr]{{\bf \boldmath (d) $ZH \rightarrow \mumutrk bb$}}
\includegraphics[height=0.20\textheight]{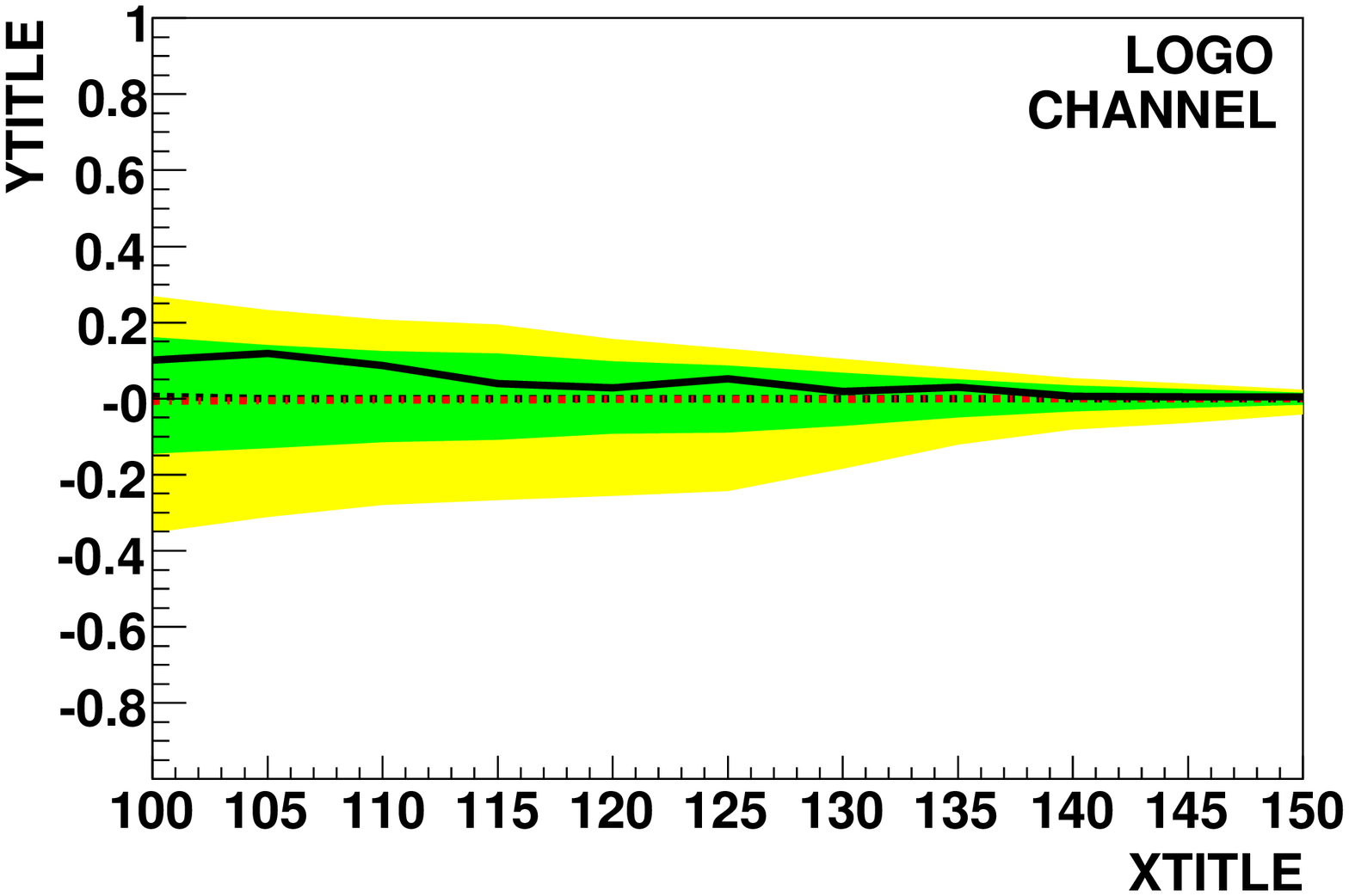}
\end{centering} \\
\end{tabular}
\caption{ Observed LLR as a function of $M_H$ for the (a) $\ee$, (b)
$\mumu$, (c) $\eeicr$, and (d) $\mumutrk$ channels.  Also shown are the
expected LLRs for the B and S+B hypotheses, together with the one and two
standard deviation (s.d.) bands about the background-only expectation.  
} \label{fig:llr_per_channel}
\end{figure*}

\begin{figure*}[htbp]
\psfrag{YTITLE}[tr][tr]{{\bf \boldmath Limit / SM}}
\psfrag{XTITLE}[tr][tr]{{\bf \boldmath $M_H$~[$\GeV$]}}
\psfrag{CHANNEL}[tl][cl]{{\bf \boldmath \mbox{D\O} 4.2 fb$^{-1}$}}
\psfrag{Observed Limit}[tl][tl]{{\bf \boldmath Observed}}
\psfrag{Expected Limit}[tl][cl]{{\bf \boldmath Expected}}
\centering
\begin{tabular}{cc}
\begin{centering}
\psfrag{LOGO}[tl][tl]{{\bf \boldmath (a) $ZH \rightarrow ee bb$}}
\includegraphics[height=0.20\textheight]{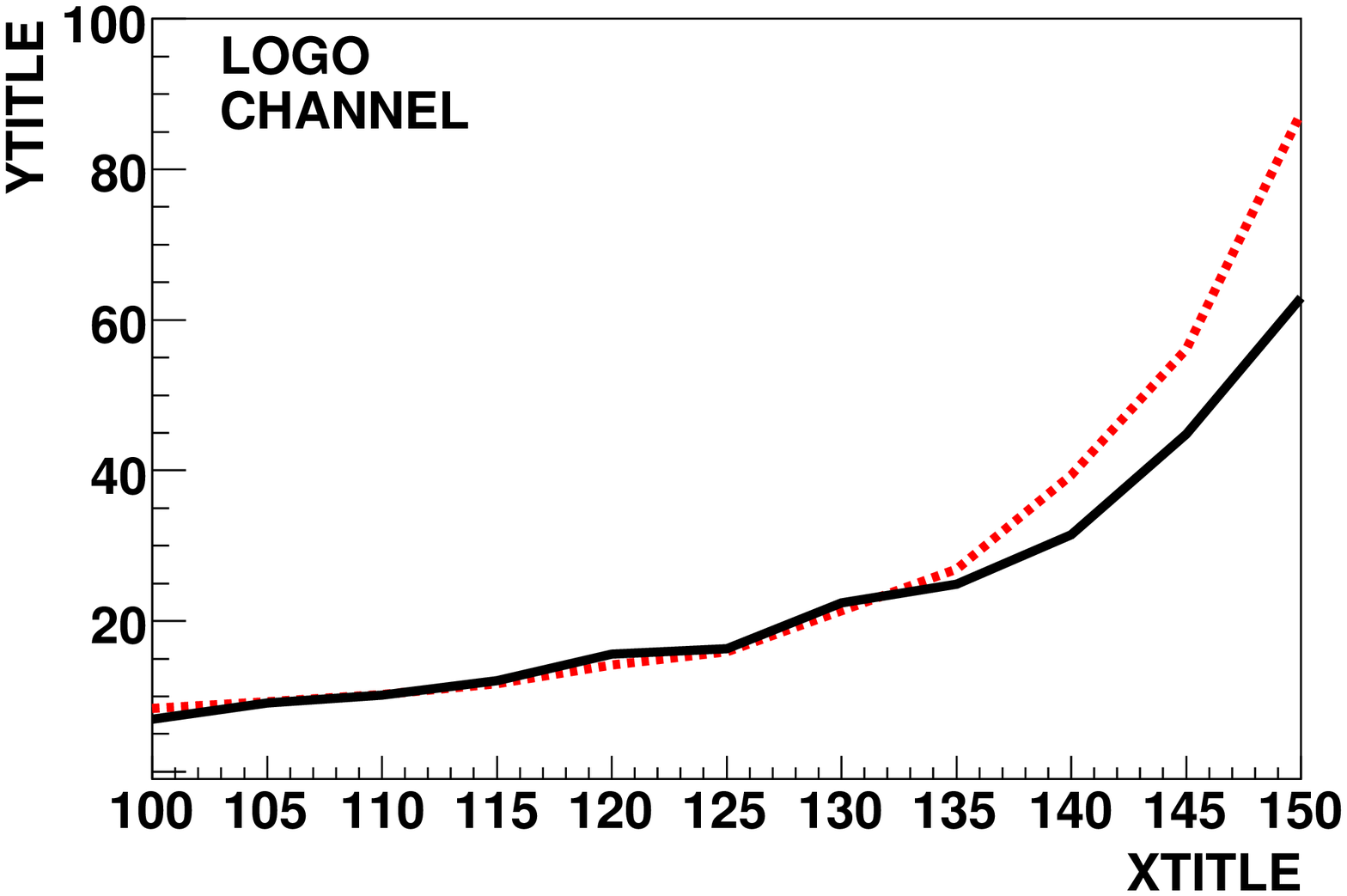}
\end{centering} &
\begin{centering}
\psfrag{LOGO}[tl][tl]{{\bf \boldmath (b) $ZH \rightarrow \mu\mu bb$}}
\includegraphics[height=0.20\textheight]{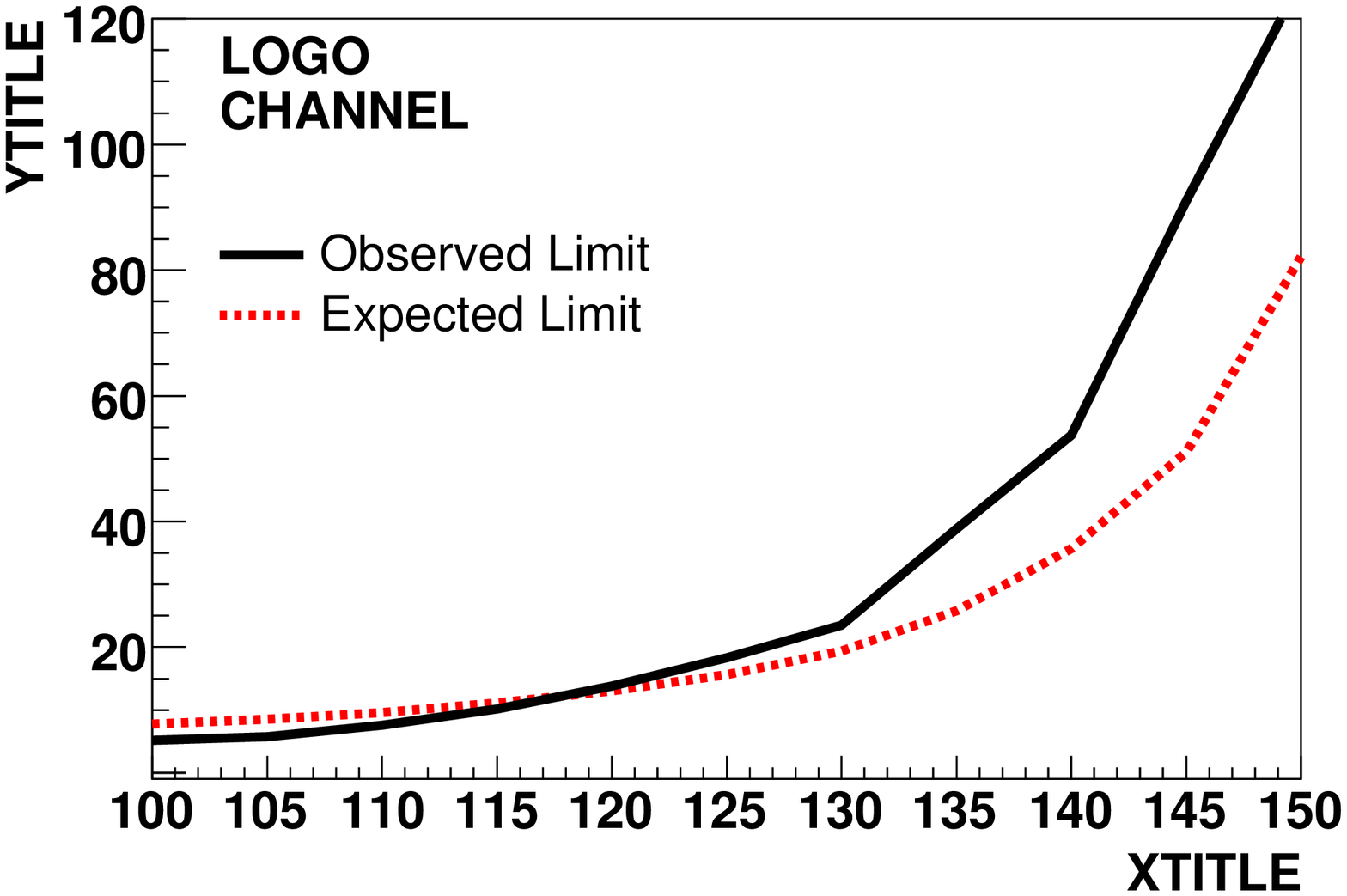}
\end{centering} \\
\begin{centering}
\psfrag{LOGO}[tl][tl]{{\bf \boldmath (c) $ZH \rightarrow \eeicr bb$}}
\includegraphics[height=0.20\textheight]{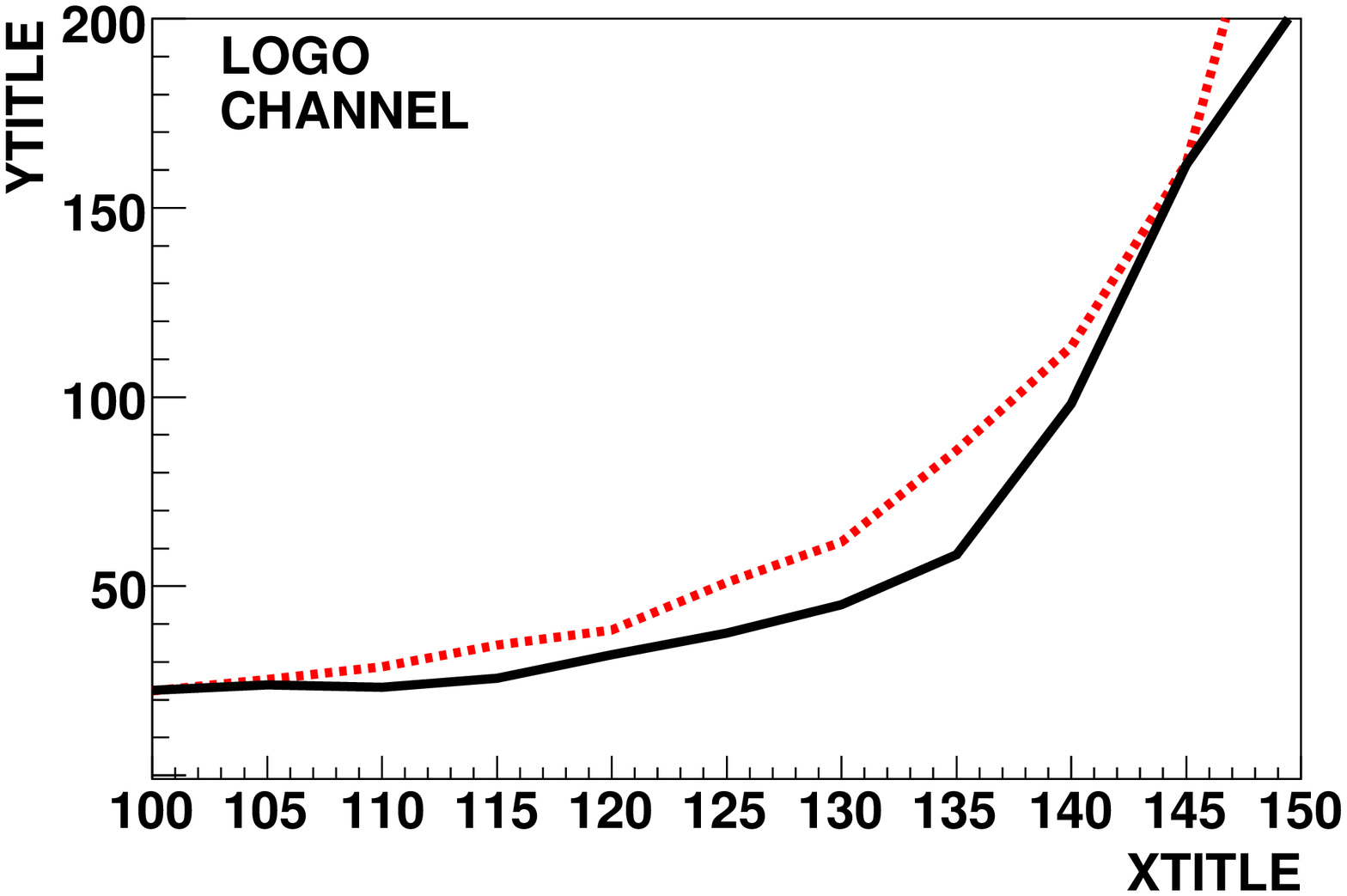}
\end{centering} &
\begin{centering}
\psfrag{LOGO}[tl][tl]{{\bf \boldmath (d) $ZH \rightarrow \mumutrk bb$}}
\includegraphics[height=0.20\textheight]{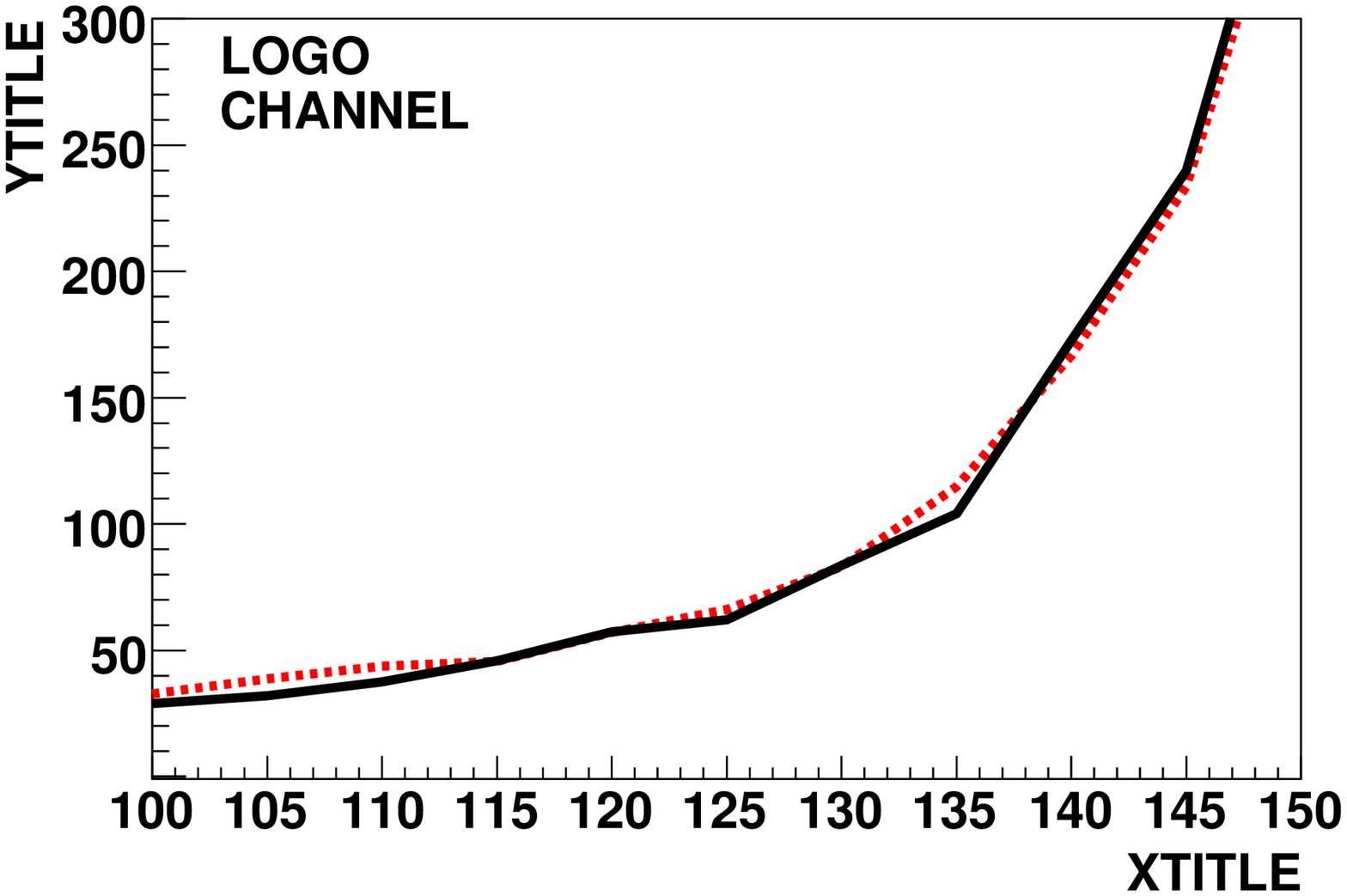}
\end{centering} \\
\end{tabular}
\caption{Expected and observed 95\% CL cross section limit divided
by the SM cross section as a function of $M_H$ for the (a) $\ee$, (b) $\mumu$, 
(c) $\eeicr$, and (d) $\mumutrk$ channels.  
}
\label{fig:lim_per_channel}
\end{figure*}


\begin{thebibliography}{99}
\bibitem{coordinates} Coordinates are relative to the proton beam direction,
with polar angle $\theta$, azimuth $\phi$, and pseudorapidity $\eta =
-\ln(\tan(\theta/2))$.
\end{thebibliography}
\end{document}
